\documentclass[%
 reprint,longbibliography,preprintnumbers,
nofootinbib,
 amsmath,amssymb,
 aps,
pre,
]{revtex4-1}
\pdfoutput=1
\usepackage{graphicx}% Include figure files
\usepackage[utf8]{inputenc}
\usepackage{flushend}
\usepackage{dcolumn}% Align table columns on decimal point
\usepackage{bm}% bold math
\usepackage{balance}
\usepackage[normalem]{ulem}

\usepackage[colorlinks = true,
            linkcolor = blue,
            urlcolor  = blue,
            citecolor =green,
            anchorcolor = blue]{hyperref}
\usepackage{verbatim}
\usepackage{color,ulem}
\usepackage[english]{babel}

\usepackage[utf8]{inputenc}
\input Starburst.fd
\newcommand*\initfamily{\usefont{U}{Starburst}{xl}{n}}\initfamily

\makeatletter 
    
\renewcommand\onecolumngrid{% <<<<<<
\do@columngrid{one}{\@ne}%
\def\set@footnotewidth{\onecolumngrid}% <<<<<<<<<<<<<<<<
\def\footnoterule{\kern-6pt\hrule width 1.5in\kern6pt}%
}

\renewcommand\twocolumngrid{% <<<<<<
        \def\footnoterule{% restore rule
        \dimen@\skip\footins\divide\dimen@\thr@@
        \kern-\dimen@\hrule width.5in\kern\dimen@}
        \do@columngrid{mlt}{\tw@}
}%

\makeatother

\newcommand{\beq}{\begin{eqnarray}}
\newcommand{\eeq}{\end{eqnarray}}
\usepackage{amsmath}
\usepackage{tikz}
\usetikzlibrary{decorations.pathmorphing}
\usetikzlibrary{shapes.misc}
\tikzset{cross/.style={cross out, draw=black, minimum size=8*(#1-\pgflinewidth), inner sep=0pt, outer sep=0pt},
cross/.default={1pt}}
\usetikzlibrary{patterns,math}
\begin{document}

\title{Shear flows in far-from-equilibrium strongly coupled fluids}

\author{\textbf{Matteo Baggioli}$^{1,2}$}%
 \email{b.matteo@sjtu.edu.cn}
 \author{\textbf{Li Li}$^{3,4,5}$}
\email{liliphy@itp.ac.cn}
 \author{\textbf{Hao-Tian Sun}$^{3,4}$}
 \email{sunhaotian@itp.ac.cn}
 \vspace{1cm}
 
\affiliation{$^{1}$Wilczek Quantum Center, School of Physics and Astronomy, Shanghai Jiao Tong University, Shanghai 200240, China}
\affiliation{$^{2}$Shanghai Research Center for Quantum Sciences, Shanghai 201315, China}
\affiliation{$^{3}$CAS Key Laboratory of Theoretical Physics, Institute of Theoretical Physics,
Chinese Academy of Sciences, Beijing 100190, China}
\affiliation{$^4$School of Physical Sciences, University of Chinese Academy of Sciences, Beijing 100049, China}
\affiliation{$^5$School of Fundamental Physics and Mathematical Sciences, Hangzhou Institute for Advanced Study, University of Chinese Academy of Sciences, Hangzhou 310024, China}

%%%%%%
\begin{abstract}
Despite the viscosity of a fluid ranges over several orders of magnitudes and is extremely sensitive to microscopic structure and molecular interactions, it has been conjectured that its (opportunely normalized) minimum displays a universal value which is experimentally approached in strongly coupled fluids such as the quark-gluon plasma. At the same time, recent findings suggest that hydrodynamics could serve as a universal attractor even when the deformation gradients are large and that dissipative transport coefficients, such as viscosity, could still display a universal behavior far-from-equilibrium. Motivated by these observations, we consider the real-time dissipative dynamics of several holographic models under large shear deformations. In all the cases considered, we observe that at late time both the viscosity-entropy density ratio and the dimensionless ratio between energy density and entropy density approach a constant value. Whenever the shear rate in units of the energy density is small at late time, these values coincide with the expectations from near equilibrium hydrodynamics. Surprisingly, even when this is not the case, and the system at late time is far from equilibrium, the viscosity-to-entropy ratio approaches a constant which decreases monotonically with the dimensionless shear rate and can be parametrically smaller than the hydrodynamic result.
\end{abstract}

\maketitle

%%%%%%%
\textit{Introduction--} Viscosity measures the resistance of a fluid to shearing motion and represents one of the most fundamental properties of liquid dynamics, whose importance ranges from biology and chemistry to cosmology and relativistic heavy-ion collisions. Despite its value spans many orders of magnitude and strongly depends on the microscopic interactions and structure, its minimum displays a certain degree of universality. This was first suggested by Purcell \cite{doi:10.1119/1.10903,trachenko2020purcell} and then emphasized in the famous Kovtun-Son-Starinets (KSS) bound \cite{Kovtun:2004de} which found already numerous experimental confirmations \cite{Sch_fer_2009,Nagle_2011,PhysRevC.78.034915,PhysRevC.84.044903} (see \cite{Cremonini:2011iq} for a review). The universal character has been also generalized to different diffusive processes \cite{Hartnoll:2014lpa} and recently confirmed in a large class of non-relativistic liquids \cite{Trachenko:2019ghg,Trachenko:2020ktm,Trachenko:2020jgr}.

From a dynamical perspective, the viscosity $\eta$ specifies the relation between the shear stress $\sigma$ and the rate of shear deformation $\dot{\gamma}$:
\begin{equation}\label{def1}
    \eta\,\equiv \frac{\sigma}{\dot{\gamma}}\,,
\end{equation}
and, for small shear rates, it can be consistently assumed to be a constant. On the contrary, when the deformation rate becomes large, this remains true only for a small subclass of systems known as \textit{Newtonian fluids}. In all other cases \cite{chhabra2010non,rivlin1948hydrodynamics}, the viscosity becomes a nonlinear function of the shear rate itself, producing a plethora of interesting and ubiquitous phenomena such as \textit{shear thicknening}, \textit{shear thinning} and many more. Familiar examples of non-Newtonian fluids are whipped cream, wall paints, blood and wet sand. In these scenarios, it is customary to define an apparent viscosity $\sigma=\eta(\dot{\gamma})\dot{\gamma}$ which converges to the near-equilibrium viscosity $\eta_0$ only in the limit $\dot{\gamma} \tau \ll 1$ (where $\tau$ is the characteristic relaxation time of the system), and that can be easily measured in rheological experiments using a viscometer.

Importantly, whenever the shear rate is large with respect
to the characteristic energy scale of the system, the fluid finds itself in a far-from-equilibrium state that cannot be described by linearized hydrodynamics, intended as the effective description of the long wave-length and late-time physics, and which a priori is not expected to reveal any degrees of universality. Interestingly, the early-time dynamics of the quark-gluon-plasma (QGP), the most perfect fluid in Nature \cite{Eskola2019,Schaefer:2014awa} is characterized by large spatial gradients \cite{PhysRevC.93.024909}. Its far-from-equilibrium nature \cite{PhysRevLett.122.122301} renders the applicability of hydrodynamics questionable (see \cite{Busza:2018rrf,Florkowski:2017olj} for reviews about the topic).

In the recent years, there has been an incredible effort in understanding whether hydrodynamics, and in which form exactly, can be still useful and applicable to describe the far-from-equilibrium dynamics of fluids~\cite{Romatschke:2017vte,Behtash:2018moe,Jaiswal:2016hex}, leading to new ideas at the edge between math and physics, such as resummed transport coefficients \cite{Denicol:2020eij}, hydrodynamics attractors \cite{Denicol:2018pak,Denicol:2019lio,Kurkela:2019set}, resurgent transseries \cite{Aniceto:2018bis} and many more. Holography \cite{Casalderrey-Solana:2011dxg,Hartnoll:2016apf,Baggioli:2019rrs} represents a very natural playground to test and explore these ideas by performing controllable computations of the real-time far-from-equilibrium dynamics of strongly coupled fluids \cite{Chesler:2008hg,Heller:2011ju,Heller:2012je,Chesler:2013lia,Romatschke:2017vte,Casalderrey-Solana:2013aba,Attems:2016tby,Ghosh:2021naw,Morales-Tejera:2020xuv,Fernandez-Pendas:2019rkh,Landsteiner:2017lwm,Grieninger:2020wsb,Ammon:2016fru}.

The current interpretation suggests that hydrodynamics may still be defined as a \textit{universal attractor} \cite{Heller:2015dha,PhysRevLett.124.102301,Du:2021fok,2018PhRvD..97c6020S,SPALINSKI2018468} and that dissipative transport coefficients, such as viscosity, can still show a universal behavior even when local gradients are large and the system is far-from-equilibrium. Following this paradigm, a natural question to ask is whether there is any remnant of the universal KSS bound far-from-equilibrium,\emph{ i.e.} for large shear rates. In particular, one would like to understand whether the dimensionless viscosity-entropy density ratio, when opportunely defined, remains constant far-from-equilibrium and whether such a value coincides or not with the KSS bound, $\eta_0/s=1/4 \pi$ (with $\hbar=k_B=1$). Here, $\eta_0$ is the value of the viscosity defined within linear response in the hydrodynamic limit and appearing as a first-order dissipative correction to the stress-energy tensor $T^{ab}$ \cite{Kovtun:2012rj}. Importantly $\eta_0$ is different from $\eta$ in Eq.\eqref{def1} for large shear rates.\\

In this Letter, we consider the real time dynamics of several bottom-up holographic models under large shear deformations, which correspond to large $N$ strongly coupled fluids far-from-equilibrium. By considering time-dependent backgrounds with finite shear deformations, we analyze the behavior of the viscosity under large gradients and explore to which extent its universal character is preserved when the system is driven far away from the equilibrium state.\\

\textit{Shear flows, viscosity and thermodynamics--} Near equilibrium, where linearized hydrodynamics applies, the viscosity can be extracted in terms of the stress tensor retarded Green's function using the standard Kubo formula \cite{Bradlyn:2012ea}:
\begin{equation}
\eta_0\,=\,-\,\lim_{\omega \,\rightarrow\, 0}\,\frac{1}{\omega}\,\textrm{Im}\,\left[\mathcal{G}^{\textrm{(R)}}_{T_{xy}T_{xy}}(\omega)\right]\,,
\label{kubo}
\end{equation}
where, for simplicity, only two spatial dimensions $(x,y)$ are considered. This procedure can be easily implemented in the holographic formalism by considering an infinitesimal gravitational shear perturbation $\delta g_{xy}\sim e^{-i \omega t}$ and computing the linear retarded response of the dual stress tensor using the holographic dictionary \cite{Son:2007vk,Policastro:2002se,Son:2002sd}. In general, whenever the shear strain rate $\dot{\gamma}$ is large, both the linear response formalism and the hydrodynamics approximation are not applicable anymore, and thus the Kubo formula in Eq.\eqref{kubo} loses its meaning. In this situation, in which the gravitational solution becomes inherently time-dependent, a more appropriate and robust way of proceeding is to compute directly the time-dependent boundary stress tensor $\sigma$ which is now a nonlinear function of $\dot{\gamma}$, and apply the more general formula presented in Eq.\eqref{def1}. As a result, the apparent viscosity $\eta$ is generally a function of both time and shear rate, thus displaying a much richer dynamics than its near equilibrium counterpart $\eta_0$. This is the procedure that will be adopted in this work.

In time-dependent holographic solutions, or equivalently in field theories out-of-equilibrium, asides from the definition of the viscosity, one must be extremely careful with the definitions of the thermodynamic quantities such as the temperature and entropy \cite{zubarev1974nonequilibrium,puglisi2017temperature,demirel2007nonequilibrium}. The standard definition of entropy, extracted using the Bekenstein-Hawking law from the area of the black-hole event horizon \cite{doi:10.1063/1.2913906}, becomes questionable. Nevertheless, there is increasing evidence (see \cite{Rougemont:2021gjm} for a detailed discussion on this point) that the correct derivation of the entropy density in out-of-equilibrium gravitational systems is through the area of the apparent horizon that is defined using local quantities \cite{Booth:2005qc}. We will therefore follow this identification. Importantly, because of the large $N$ limit, the effects of hydrodynamic fluctuations which are known to spoil the late time behavior of two dimensional fluids \cite{PhysRevA.16.732} are neglected in our computations. We do expect the picture emerging from our analysis to be, at least qualitatively, similar to that in higher dimensions where the effects of fluctuations become irrelevant.\\

\textit{Out of equilibrium steady states--}  We consider three different bottom-up models in asymptotically AdS$_4$ spacetime. The first setup is the standard Einstein-Maxwell (EM) action whose field theory dual represents a $(2+1)$ dimensional strongly coupled relativistic charged fluid with a global $U(1)$ symmetry \cite{Hartnoll:2009sz}. The second framework contains a different deformation which introduces a non-trivial elastic bulk modulus in the neutral dual field theory (see \cite{Alberte:2015isw,Alberte:2016xja,Baggioli:2016rdj,Andrade:2019zey,Baggioli:2019abx,Baggioli:2021xuv} for more details). Finally, in the third one, conformal symmetry is explicitly broken by a scalar deformation and the corresponding boundary stress-tensor is not traceless anymore \cite{Li:2020spf}. Near equilibrium (in the hydrodynamic limit), all the dual fluid field theories considered display a universal value for the viscosity-to-entropy density ratio which saturates the KSS bound $\eta_0/s\,=\,1/4\pi$. All the additional details about the models can be found in the appendices.
\begin{figure*}[ht]
		\begin{center}
			\includegraphics[width=0.32\textwidth]{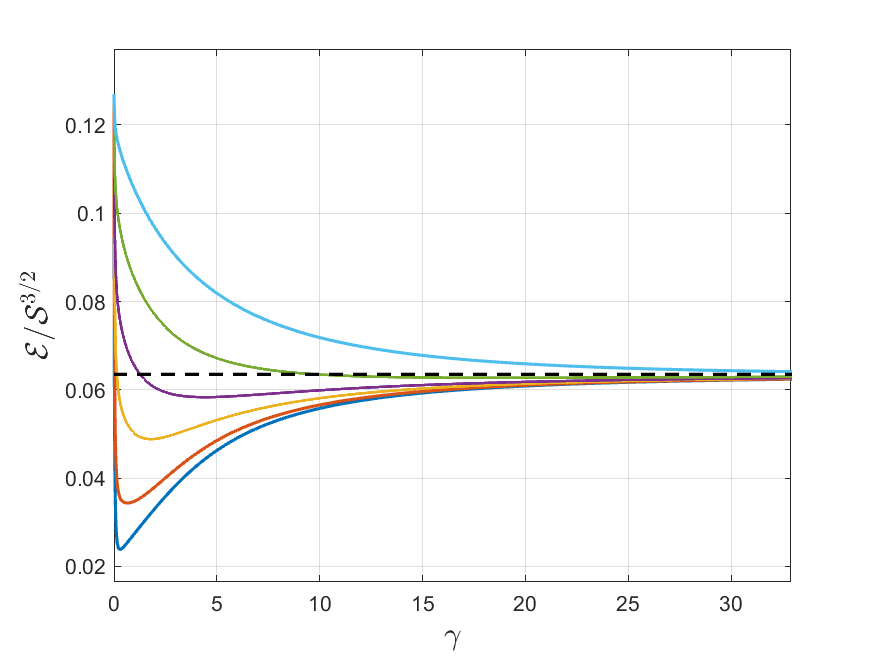}
			\includegraphics[width=0.32\textwidth]{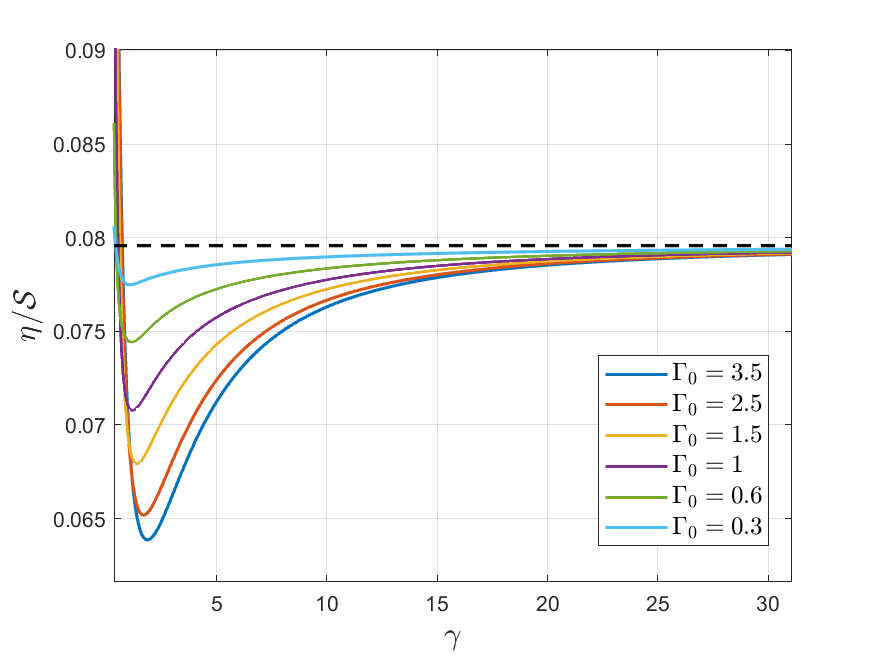}
				\includegraphics[width=0.32\textwidth]{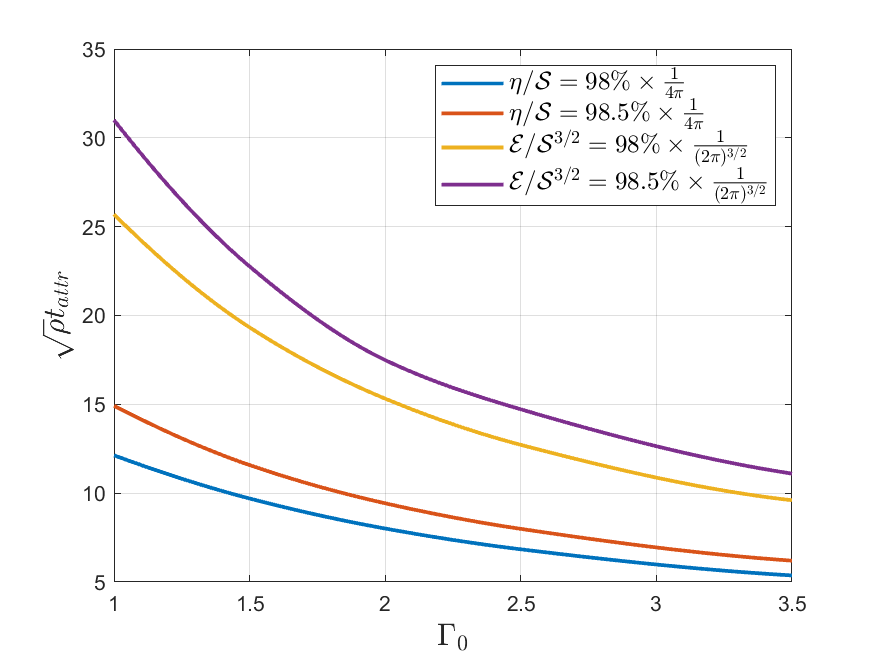}
			\caption{ \textbf{Left: } $\mathcal{E}/\mathcal{S}^{3/2}$ as a function of $\gamma(t)$, where the dash line is $1/(2\pi)^{3/2}$. \textbf{Center: } $\eta/\mathcal{S}$ as a function of $\gamma(t)$, where the dash line is $1/4\pi$. \textbf{Right: } The dimensionless attracting times, defined using the $1.5\%$ and $2\%$ rules, for both the viscosity and energy density ratios as a function of the normalized shear rate.}
			\label{fig:1}
		\end{center}
\end{figure*}

In order to drive our holographic systems far from equilibrium, we introduce a time dependent source for the metric component $g_{xy}$. From the dual field theory point of view, that corresponds to deform our fluid with a time dependent shear strain $\gamma(t)$. In the main text, we will focus on the EM model and two benchmark shear rates. Different holographic models (including a non conformal one) and different shear strain deformations are discussed in Appendices \ref{s4}, \ref{snonconf} and give equivalent results. We construct numerically the far-from-equilibrium geometry which allows us to read all the physical observables of the dual field theory. Importantly, because the boundary field theory geometry is not of Minkowski type, extra-care needs to be taken in order to define the boundary shear stress $\sigma(t)$ and the shear strain $\gamma(t)$. All the details of our computations can be found in Appendices \ref{s1}-\ref{s5}.

Let us first consider the EM model in presence of a constant shear rate $\dot{\gamma}\equiv \gamma_0$. The time dependent dynamics (see Fig.~\ref{fig:1old}) strongly depends on the dimensionless shear rate $\Gamma_0\equiv \gamma_0/\sqrt{\rho}$, where $\rho$ is the charge density of the dual field theory. For small $\Gamma_0$, the time evolution is slow and all quantities slowly grow in time. Whenever the gradients are large, all quantities rapidly increase and deviate from their initial values. The viscosity and the entropy density grow quadratically in time, $\sim t^2$, while the energy density grows like $\sim t^3$. More in general, we find that, independently of the shear deformation used, the behavior of the energy density is constrained by one of the Einstein's equations to follow:
\begin{equation}
\mathcal{E}(t)-\mathcal{E}_{\mathrm{ini}}=\int_{t_{\mathrm{ini}}}^t \dot{\gamma}(\tau)\sigma(\tau) d\tau=\int_{t_{\mathrm{ini}}}^t \eta(\tau)\dot{\gamma}(\tau)^2 d\tau\,, \label{ee}
\end{equation}
from which the cubic scaling mentioned above can be immediately derived. Here, the subscript \textit{ini} refers to the initial configuration on top of which the shear deformation is switched on. Eq.\eqref{ee}, which is obtained analytically from the gravitational setup (see Appendix~\ref{s3}), corresponds exactly to the expression for the dissipated energy in a viscoelastic system under deformations \cite{nla.cat-vn1732309}. Let us emphasize that, in order to avoid infinite gradients during the introduction of the constant strain rate, the time dependent boundary strain $\gamma(t)$ presents an initial activation window which is responsible for the non-monotonic oscillations observed in all our results at early time. We explicitly checked that the form of the activation function does not affect our results.

From a physical perspective, we can notice that the apparent viscosity $\eta$ grows with the shear rate $\Gamma_0$. In the context of rheology, this behavior is denoted as \textit{shear thickening} and it is typical of dilatant fluids, such as blood, ketchup and peanut butter. These results are compatible with those of \cite{Baggioli:2019mck}, found using oscillatory strain methods.

The dynamics of the two dimensionless ratios $\mathcal{E}/\mathcal{S}^{3/2}$, $\eta/\mathcal{S}$, with $\mathcal{S}$ the entropy density, is shown in Fig.\ref{fig:1}. The time evolution profiles present a transient regime which is highly sensitive to the initial conditions and to the rate of shear deformation $\Gamma_0$. In that window, none of the quantities seem to follow any specific trend and the UV microscopic details are dominating the dynamics. The deviations from the initial value can reach up to $50\%$ and are larger by increasing the rate of deformation $\Gamma_0$. Moreover, the viscosity-to-entropy density ratio clearly violates the KSS bound in agreement with the results of \cite{Wondrak:2020tzt}. Nevertheless, after a certain time, which we label as the \textit{attracting time} (see below for a more formal definition), we observe that both quantities approach a constant value which is given exactly by the close-to-equilibrium hydrodynamics expectations. The same universal behavior is observed in the other holographic models considered and, surprisingly, also for a non-conformal model (see Appendix \ref{snonconf}) and for deformations with a non-constant shear rates such as $\gamma(t)\sim \sqrt{t},t^2,t^4$ (see Appendix \ref{s4}).

To continue, let us define the \textit{attracting time} $t_{\mathrm{attr}}$ as the value at which the dimensionless quantities reach their final late-time values. Pragmatically, we decided to use two different criteria coinciding with the $98\%$ and the $98.5\%$ of the final values.  We show this quantity for both the viscosity and the entropy density as a function of the normalized shear rate in the right panel of Fig.~\ref{fig:1}. Interestingly, we find that the larger the shear rate, the faster the time evolution reaches the universal attractor.

\begin{figure}
	   \centering
	    \includegraphics[width=0.35\textwidth]{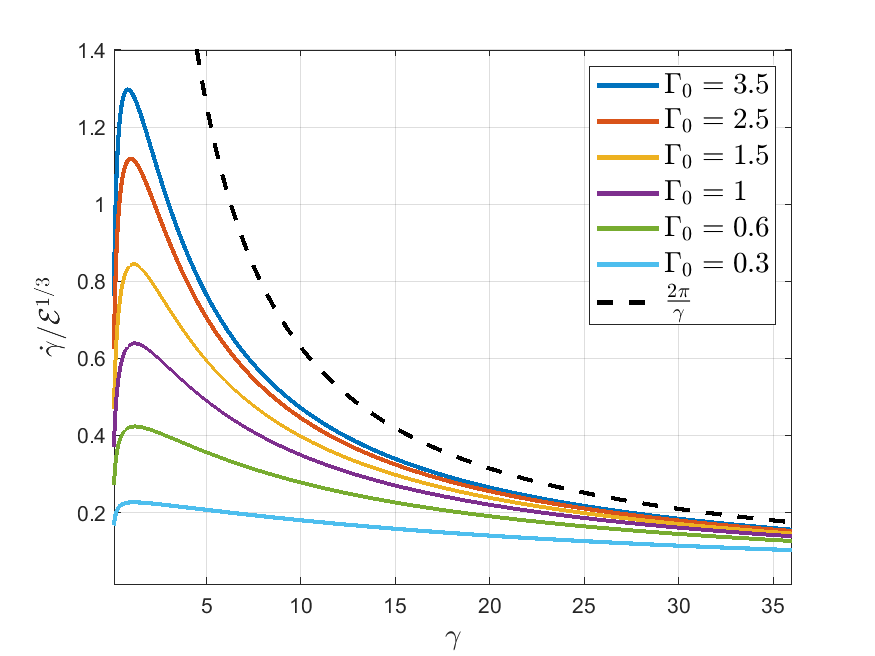}
	    
	    \vspace{0.1cm}
	    
	    \includegraphics[width=0.35\textwidth]{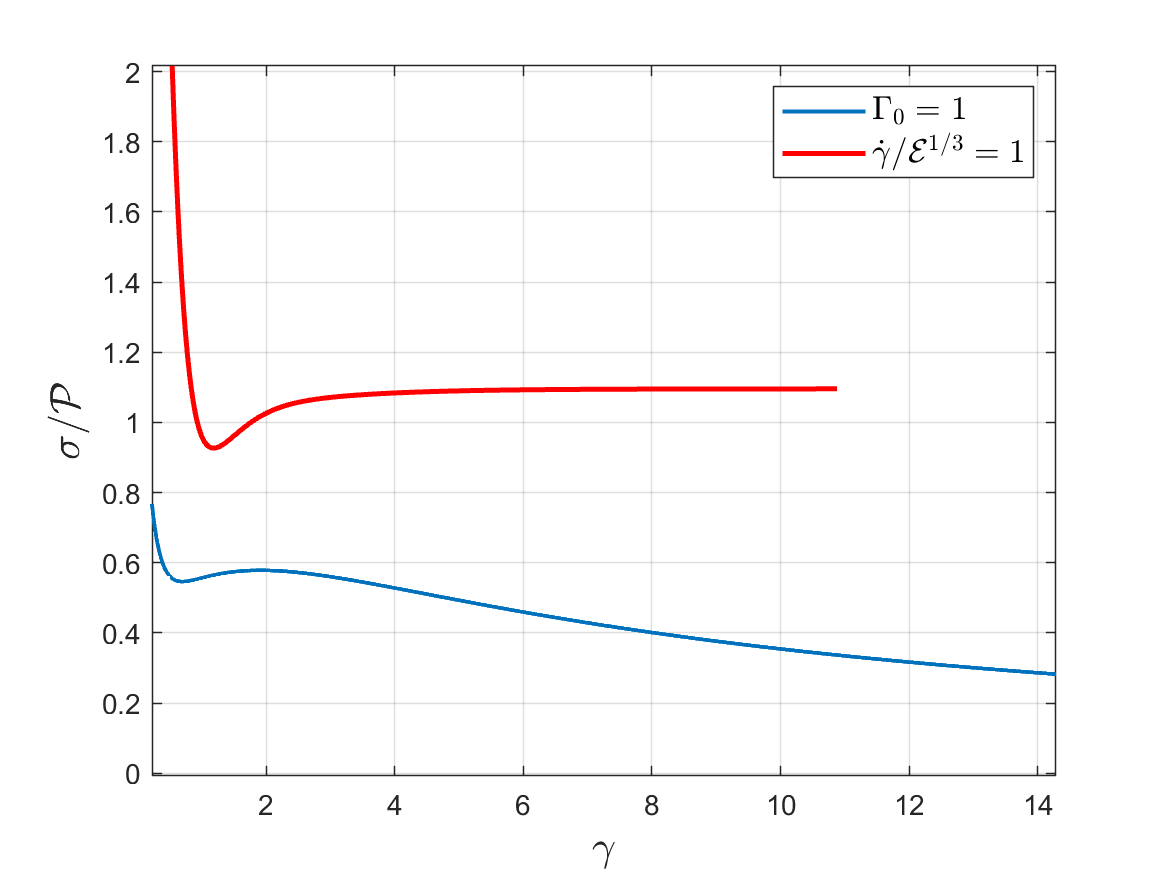}
	    
	    \caption{\textbf{Top: }The dimensionless combination $\dot{\gamma}/\mathcal{E}^{1/3}$ as a function of $\gamma$ for different shear rates. The dashed line is the attractor function $2 \pi/\gamma$ approached by all curves for $\gamma \gg 1$. \textbf{Bottom: }The pressure anisotropy $\sigma/\mathcal{P}$ as a function of $\gamma(t)$ for constant $\Gamma_0=1$ (blue) and at constant $\dot{\gamma}/\mathcal{E}^{1/3}=1$ (red) in the EM model.}
	    \label{fig:2}
\end{figure}

In order to understand this emerging universality, in Fig.~\ref{fig:2}, we plot the dimensionless ratio between the shear rate and the energy density of the system. We find that for large deformations, $\gamma \gg 1$, such a ratio goes to zero following a universal law $\dot{\gamma}/\mathcal{E}^{1/3}=2 \pi/\gamma$. This numerical observation implies the existence of a universal attractor solution in the regime $\gamma(t)=\gamma_0 t \gg 1$ given by
\begin{equation}
    \mathcal{E}= \left(\frac{\gamma_0}{2 \pi}\right)^3\,\gamma^3\,,\quad \eta=\frac{\gamma_0^2}{2\,(2 \pi)^2}\,\gamma^2\,,\quad  \mathcal{S}=\frac{\gamma_0^2}{2 \pi}\,\gamma^2\,,\label{attra}
\end{equation}
which is consistent with the data presented in Fig.~\ref{fig:1}.\\

Note that, in far-from-equilibrium systems, a local rest frame might be absent \cite{Arnold:2014jva}. In all the cases considered, a local rest frame exists (see appendix \ref{s2}) and therefore the notion of hydrodynamic attractor is always well defined \cite{Arnold:2014jva}.\\

\textit{A far-from-equilibrium case--} In the setups considered so far, the shear rate in units of the characteristic energy scale of the system becomes small at late time (see top panel of Fig.\ref{fig:2}). This suggests the presence of a late time steady state which is effectively in equilibrium. Therefore, it is perhaps not surprising that the near equilibrium hydrodynamic results apply in these situations.

\begin{figure}
	 \centering
	  \includegraphics[width=0.35\textwidth]{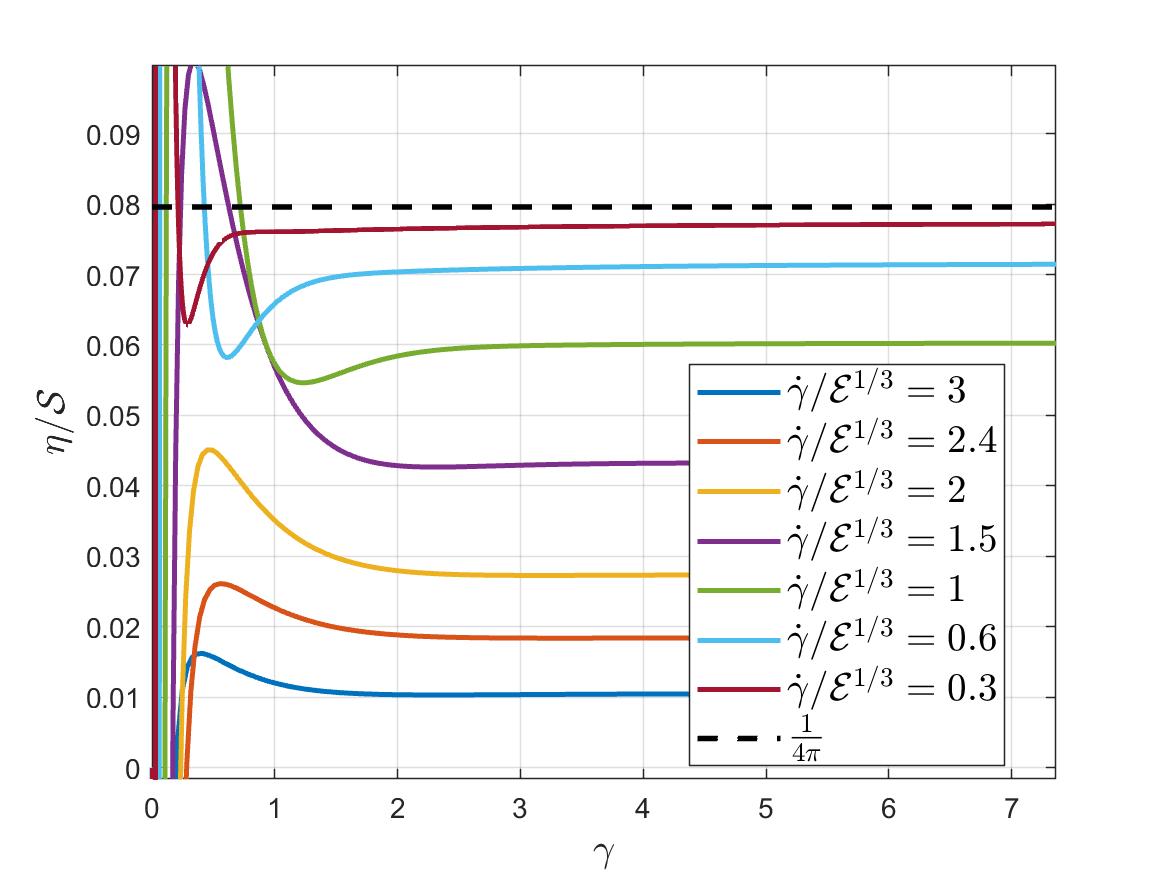}
	    
	    \vspace{0.1cm}
	    
	    \includegraphics[width=0.35\textwidth]{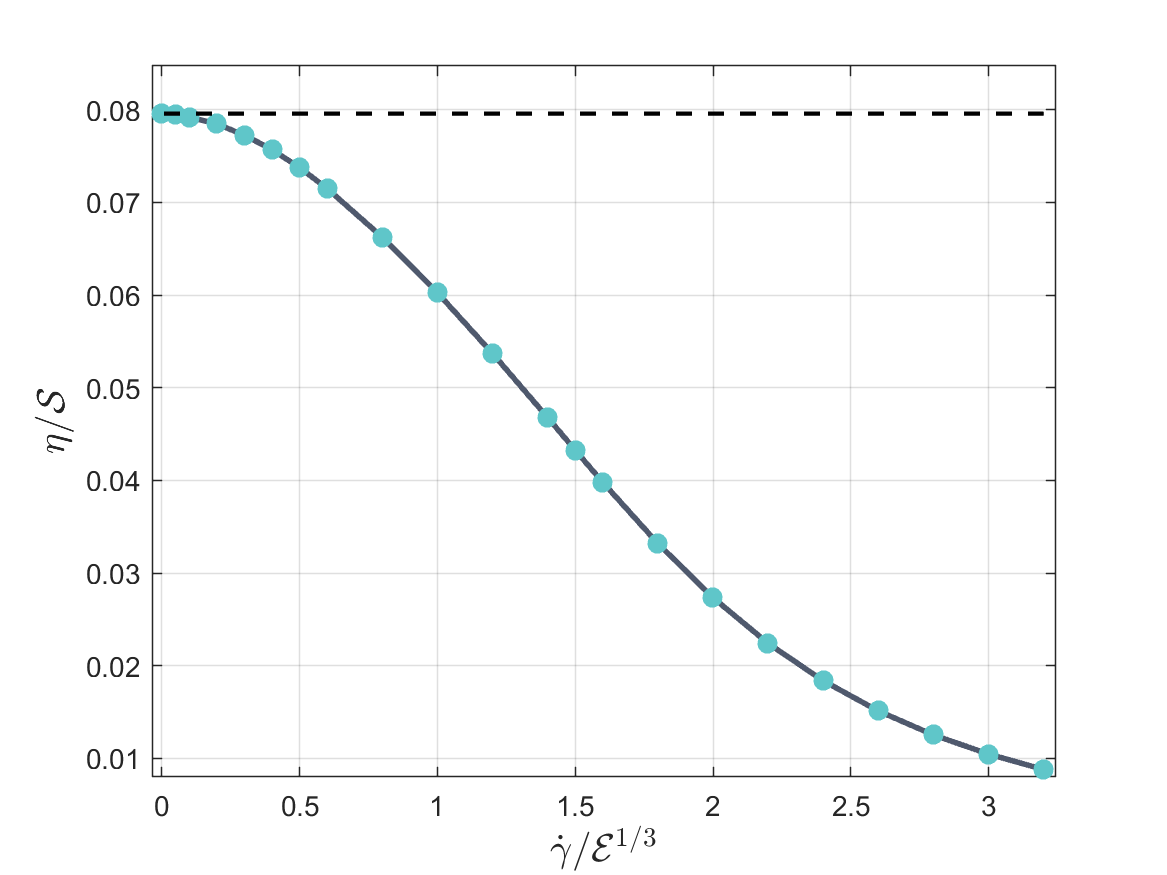}
	    \caption{The out of equilibrium steady state. {\bf Top: }$\eta/s$ as a function of $\gamma$ for different values of the dimensionless gradient $\dot{\gamma}/\mathcal{E}^{1/3}$. {\bf Bottom: } The value of the $\eta/s$ ratio at late time as a function of the dimensionless shear rate $\dot{\gamma}/\mathcal{E}^{1/3}$.}
	    \label{fig:3}
\end{figure}

By considering a different time-dependent shear rate (see details in Appendix \ref{sfar}), we are able to keep the energy-normalized shear rate fixed, and arbitrarily large, at late time. In this case, the system never reaches an effective equilibrium state in which all physical quantities are time dependent but the gradients are small compared to the characteristic energy scale. This distinction is confirmed by the analysis of the pressure anisotropy  $\Delta \mathcal{P}\equiv \sigma/\mathcal{P}$ at late time, $\gamma \rightarrow \infty$ (bottom panel of Fig.\ref{fig:2}). Whether for the previous cases (blue curve) $\Delta \mathcal{P} \rightarrow 0$ at late times, in this new setup (red curve) it approaches an $\mathcal{O}(1)$ constant signaling the far-from-equilibrium nature of the late time state. Interestingly, even in this far-from-equilibrium situation, both the viscosity to entropy ratio and the dimensionless energy density approach a constant value in the late time steady state, as shown in the top panel of Fig.\ref{fig:3}. Nevertheless, this value does not coincide anymore with the near-equilibrium expectation, \emph{e.g.} $1/4\pi$ for the viscosity. On the contrary, as shown in the bottom panel of Fig.\ref{fig:3}, the out of equilibrium value for $\eta/\mathcal{S}$ decays monotonically with the strength of the shear gradient and seems to approach zero for extremely large values of the shear rate. This indicates that, even far from equilibrium, the system is described by an effective hydrodynamic steady state whose transport properties are nevertheless parametrically different with respect to its near-equilibrium counterpart. Our results are consistent with the findings of \cite{romatschke2017relativistic} for the highly symmetric Bjorken flow in which the viscosity out of equilibrium was found to be parametrically smaller than the equilibrium value.\\

\textit{Outlook--} In summary, we have performed an extensive time-dependent numerical analysis of several bottom-up (conformal and not) holographic models driven away from equilibrium by different shear rates. Whenever the shear rate in units of the energy density becomes small at late time, our results reveal the emergence of a universal attracting behavior, encoded in the simple solution \eqref{attra}, on which both the dimensionless viscosity-entropy and energy-entropy ratios reach a constant value which coincides exactly with the prediction of hydrodynamics naively valid only near equilibrium. Importantly, we prove that this behavior persists also when conformal symmetry is abandoned (see \emph{e.g.} \cite{romatschke2017relativistic,Chattopadhyay:2021ive,Chen:2021wwh} for similar discussions).

In the second scenario, in which the shear rate is kept constant and large in units of the energy density, we still observe a late time steady state on which the $\eta/\mathcal{S}$ ratio takes a constant value. Nevertheless, we find that such a value does not coincide anymore with the near equilibrium hydrodynamic result $1/4\pi$ but it rather decreases monotonically with the strength of the gradients, becoming parametrically smaller than the KSS bound.

Our results provide another case in favor of ``the unreasonable effectiveness"  \cite{hong,Noronha-Hostler:2015wft} of a nonlinearly renormalized version of hydrodynamics out of equilibrium and might have important consequences not only on the out-of-equilibrium dynamics in heavy-ion collisions and QGP \cite{Heller:2016gbp,Florkowski:2017olj,Romatschke:2009im,Jaiswal:2016hex} but also on the understanding and characterization of rheological response of complex fluids \cite{Andrade:2019zey,Baggioli:2019mck,Pan:2021cux} using the holographic tool \cite{DeWolfe:2013cua}. As a roadmap for the future, it would be interesting to understand whether the existence of the observed far from equilibrium steady states is universal and whether the corresponding transport properties can be related (probably in a highly nonlocal way) to the near equilibrium counterparts (\textit{e.g.} by promoting the transport coefficients to be nonlinear functions of the deformations or by resumming the nonlinear effects).

\subsection*{Acknowledgments} 
We thank A.~Buchel, S.~Grieninger, A.~Soloviev, J.~Noronha, L.~Noirez, W.~van der Schee, I.~Aniceto and A.~Zaccone for useful comments and suggestions. M.B. acknowledges the support of the Shanghai Municipal Science and Technology Major Project (Grant No.2019SHZDZX01). L.L. acknowledges the support from the National Natural Science Foundation of China Grants No.12122513, No.12075298, No.11991052, No.12047503, and from the Key Research Program of the Chinese Academy of Sciences (CAS) Grant NO. XDPB15 and the CAS Project for Young Scientists in Basic Research YSBR-006.
 
\bibliographystyle{apsrev4-1}
\bibliography{references}
\newpage
\onecolumngrid
\appendix 
\clearpage
\renewcommand\thefigure{S\arabic{figure}}    
\setcounter{figure}{0} 
\renewcommand{\theequation}{S\arabic{equation}}
\setcounter{equation}{0}
\renewcommand{\thesubsection}{SI\arabic{subsection}}
\section*{Supplementary Information}
\subsection{The holographic conformal setup}\label{s1}
We consider a $4$-dimensional holographic setup defined by the following action principle.
\begin{equation}\label{action}
S\,=\frac{1}{2\kappa_N^2}\,\int d^4x \sqrt{-g}
\left[\mathcal{R}-2 \Lambda-c_1\,\frac{1}{4}F_{\mu\nu}F^{\mu\nu}-c_2\, Z^2\right]\,.
\end{equation}
Here $\kappa_N^2=8\pi G_N$ is the 4D gravitational Newton’s constant, $\ell$ is the asymptotic $AdS$ radius, $\Lambda=-3/\ell^2$ is a negative cosmological constant and $F=d A$ the usual Maxwell's field strength. We set $\ell=1$. Moreover, we define:
\begin{equation}
    Z=\mathrm{Det} \left[\partial_\mu \psi^I \partial^\mu \psi^J\right],\quad I,J=\{x,y\}\,,
    \label{sss}
\end{equation}
 in terms of a doublet of neutral scalars whose profile is chosen to be linear in the spatial coordinates of the boundary field theory.
 
The action parameters $c_n$ are conveniently chosen to separate two different classes of models. The first one ($c_2=0,c_1=1$) is the standard Einstein-Maxwell (EM) action
\begin{equation}\label{model1}
S_1\,=\frac{1}{2\kappa_N^2}\,\int d^4x \sqrt{-g}
\left[\mathcal{R}-2 \Lambda-\frac{1}{4}F_{\mu\nu}F^{\mu\nu}\right]\,,
\end{equation}
which corresponds to a (2+1)-dimensional relativistic (conformal) charged fluid. The second model ($c_1=0,c_2=1$) 
\begin{equation}\label{model2}
S_2\,=\frac{1}{2\kappa_N^2}\,\int d^4x \sqrt{-g}
\left[\mathcal{R}-2 \Lambda- Z^2\right]\,,
\end{equation}
corresponds to a neutral state. The operators dual to the $\psi^I$ bulk fields break spontaneously translational invariance in the dual field theory. Nevertheless, $Z$ is invariant under volume-preserving internal diffeomorphisms in the space of the $\psi^I$. In accordance with the effective field theory constructions of \cite{Nicolis:2015sra}, this implies that such bulk deformation does not introduce any shear rigidity but only a finite bulk modulus \cite{Alberte:2015isw} (which obviously realistic liquids possess as well). In other words, differently from similar systems breaking spontaneously translations \cite{Alberte:2017oqx}, the phase described by the dual field theory in this case remains fluid (see \cite{Baggioli:2019jcm} for a review about the distinctions between solids and fluids). Moreover, as shown in \cite{Alberte:2016xja}, these models, in the near-equilibrium limit, saturate the KSS bound for the viscosity-to-entropy density ratio. For more details about this construction we refer to \cite{Baggioli:2016rdj,Baggioli:2021xuv}.

In order to introduce non-trivial shear deformations in the boundary field theory, we consider the following time dependent metric in Eddington-Finkelstein(EF) coordinates with $g_{xy}\neq0$ :
\begin{equation}\label{appansatz}
ds^2=-B(u,t)dt^2-\frac{2 \,du\, dt}{u^2}+\Sigma(u,t)^2\left[ \cosh{\left(H(u,t)\right)} (dx^2+dy^2)+2\sinh{\left(H(u,t)\right)} dx \,dy \right]\,,
\end{equation}
together with the solutions for the matter fields
\begin{equation}
    \psi^x=\alpha\, x,\quad \psi^y=\alpha\, y,\quad A_\mu dx^\mu=\Phi(u,t) dt\,.
\end{equation}
The constant $\alpha$ parametrizes the size of the bulk modulus in the dual field theory while the gauge field $A_\mu$ introduces a finite charge density $\rho$ which is identified with its subleading term in the boundary expansion (see details below).\\
In our notations, the asymptotical AdS boundary is located at $u=0$. Instead of considering the event horizon that depends
on the entire history of the geometry, in the non-equilibrium case, an appropriate interior boundary is the apparent horizon which is the outermost trapped null surface located at $u=u_A$ and formed behind the event horizon. The details about its identification in the numerical routine are provided in Appendix \ref{s5}.\\

The bulk equations of motion for our system are given by
\begin{eqnarray}
&&\Sigma{}^{''} +\frac{2}{u}\Sigma{}^{'}+\frac{{H{}^{'}}^2 }{4}\Sigma=0\,,\\
&&(d_+\Sigma){}^{'} +\frac{\Sigma{}^{'} }{\Sigma } {d_+\Sigma}=-\frac{3\Sigma}{2 u^2}+\frac{1}{8}u^2\Sigma\Phi'^2+\frac{\alpha^8}{4u^2\Sigma^7}\,,\\
&&(d_+H){}^{'} +\frac{ \Sigma^{'} }{\Sigma}d_+H=-\frac{H{}^{'} {d_+\Sigma} }{\Sigma }\,,\\
&&B{}^{''}+\frac{2 }{u}B{}^{'}= \frac{H{}^{'}{d_+H}} {u^2}-\frac{4\Sigma{}^{'}{d_+\Sigma}}{u^2 \Sigma^2}+\frac{4\alpha^8}{u^4\Sigma^8}+\Phi'^2\,,\\
&&4{d_+^2\Sigma} +2 u^2 B{}^{'}d_+\Sigma+ {(d_+H)}^2 \Sigma =0\,,\label{eq:boundaryB}\\
&&\frac{\mathrm{d} \Phi'}{\mathrm{d} t}=-\frac{2d_+\Sigma+u^2B\Sigma'}{\Sigma}\Phi',\\
&&\Phi''+\frac{2\Phi'(\Sigma+u\Sigma')}{\Sigma u}=0.
\end{eqnarray}
The equations of motion for the scalars $\psi^{I}$ are trivially satisfied.
In the expressions above, we have introduced the directional derivative $d_+\mathcal{F}:= \dot{\mathcal{F}}-\frac{u^2 B}{2}\mathcal{F}^{'}$, with prime denoting the derivative with respect to radial coordinate $u$ and dot the derivative with respect to time $t$. We have also chosen $c_1=c_2=1$ for later convenience. The Einstein-Maxwell model can be recovered by setting $\alpha=0$ while the second one by setting $\Phi=0$.
For the radial integration of the equations of motion above, we consider the bulk region spanning from the AdS boundary $u=0$ to the apparent horizon located at $u_A$. The boundary asymptotics for the different bulk fields are found to be
\begin{equation}\label{bdyexpansion}
\begin{split}
&B=\frac{1}{u^2}\left[1+\frac{2(s_1-\dot{s}_0)}{s_0}u+\left( \frac{s_1^2}{s_0^2}-\frac{2\dot{s}_1}{s_0}-\frac{3\dot{h}_0^2}{4} \right)u^2+b_3 u^3+\mathcal{O}(u^4)\right],\\
&\Sigma=\frac{1}{u}\left[s_0+s_1 u-\frac{s_0\dot{h}_0^2}{8}u^2+\frac{s_1 \dot{h}_0^2}{8} u^3+\mathcal{O}(u^3)\right],\\
&H=h_0+\dot{h}_0 u-\frac{s_1 \dot{h}_0}{s_0}u^2+h_3 u^3+\mathcal{O}(u^4)\,,\\
&\Phi=\phi_0(t)+\phi_1(t)u+\mathcal{O}(u^2)\,,
\end{split}
\end{equation}
together with the constraints:
\begin{equation}\label{eq:constraint}
\begin{split}
\dot{b}_3+\frac{3 b_3 \dot{s}_0}{s_0}-\dot{h}_0^2 \left(\frac{\dot{s}_0^2-3 s_1^2}{2 s_0^2}+\frac{\ddot{s}_0}{2 s_0}\right)-\frac{3 \dot{h}_0 \ddot{h}_0 \dot{s}_0}{2 s_0}
+\frac{3}{8} \dot{h}_0^4-\frac{3}{2} h_3 \dot{h}_0-\frac{1}{2} \dddot{h}_0 \dot{h}_0=0\,,\\
\dot{\phi}_1 s_0+2\phi_1\dot{s}_0=0\,.
\end{split}
\end{equation}
Without loss of generality, we fix the apparent horizon at $u_A=1$ using the residual symmetry and we keep $s_1$ as a dynamical parameter.\,\footnote{Note that the EF coordinates have a residual diffeomorphism invariance $u\rightarrow \tilde{u}=\frac{u}{1+\lambda(t)u}$  with $\lambda(t)$ an arbitrary function of time, under which $s_1\rightarrow\tilde{s}_1=s_1-\lambda(t)s_0$.}

The renormalized action is obtained by adding the necessary counterterms as well as the standard Gibbons-Hawking term 
\begin{equation}
S_{\text{ren}}=S+\frac{1}{2\kappa_N^2}\int_\partial\sqrt{-h}\left(2\mathcal{K}+4-\mathcal{R}^h\right)\,,
\end{equation}
where $h_{ab}$ is the induced metric on the boundary, $\mathcal{R}^h$ the associated scalar curvature and $\mathcal{K}$ the trace of the extrinsic curvature, given by  $\mathcal{K}_{\mu\nu}={h_\mu}^\lambda {h_\nu}^\sigma\nabla_\lambda n_\sigma$ with $n$ the outward pointing unit normal to the boundary. The boundary stress tensor is then given by
\begin{equation}\label{Htensor}
T_{ab}=\lim_{u\rightarrow 0}-\frac{2}{u\sqrt{-h}}\frac{\delta S_{\text{ren}}}{\delta h^{ab}}=\frac{1}{2\kappa_N^2}\lim_{u\rightarrow 0}\frac{2}{u}\,\left(\mathcal{K} \,h_{ab}-\mathcal{K}_{ab}-2h_{ab}+G^h_{ab}\right)\,,
\end{equation}
with $G^h_{ab}$ being the Einstein tensor in terms of the boundary metric $h_{ab}.$
Finally, the $U(1)$ current can be defined as
\begin{equation}
J^a=\lim_{u\rightarrow 0}\frac{1}{u^3\sqrt{-h}}\frac{\delta S_{\text{ren}}}{\delta A_a}=\frac{1}{2\kappa_N^2}\lim_{u\rightarrow 0}\frac{1}{u^3}n_\nu F^{a\nu}\,.
\end{equation}
Latin indices run through boundary directions only, \emph{i.e.} $a=t, x, y$; Greek indices run over the bulk spacetime dimensions, \emph{i.e.} $\mu=u,t, x, y$.  

The non-vanishing components of $T_{ab}$ and $J^a$ are given by
\begin{eqnarray}
&&\mathcal{E}\equiv T_{tt} ={ -b_3}\,,\label{stress0}\\
&&T_{xx} =T_{yy}={ -\frac{1}{2}b_3s_0^2\cosh{(h_0)}} - \frac{3}{2}s_1^2\dot{h}_0\sinh{(h_0)}+\frac{3}{2}h_3 s_0^2\sinh({h_0)}\nonumber\\
&&-\frac{s_0^2}{8}\sinh{(h_0)}\left(3\dot{h}_0^3-4\dddot{h}_0-\frac{4\dot{h}_0\dot{s}_0^2}{s_0^2}-\frac{4(\dot{h}_0\ddot{s}_0+3\ddot{h}_0\dot{s}_0)}{s_0}\right)\,,\\
&&T_{xy} =T_{yx}={ -\frac{1}{2}b_3s_0^2\sinh{(h_0)}} - \frac{3}{2}s_1^2\dot{h}_0\cosh{(h_0)}+\frac{3}{2}h_3 s_0^2\cosh({h_0)}\nonumber\\
&&-\frac{s_0^2}{8}\cosh{(h_0)}\left(3\dot{h}_0^3-4\dddot{h}_0-\frac{4\dot{h}_0\dot{s}_0^2}{s_0^2}-\frac{4(\dot{h}_0\ddot{s}_0+3\ddot{h}_0\dot{s}_0)}{s_0}\right)\,,\label{stress1}\\
&&\rho=J^t=-\phi_1/2\,,
\end{eqnarray}
where $\mathcal{E}$ and $\rho$ indicate, respectively, the energy density and the charge density of the dual field theory. Here, we have chosen $\kappa_N^2=1$.
The boundary metric, in which the dual field theory lives, reads
\begin{equation}\label{Fmetric}
\begin{split}
d s^2=&\gamma_{ab}dx^a dx^b
=-d t^2+s_0(t)^2[\cosh{(h_0(t))}(dx^2+dy^2)+2\sinh{(h_0(t))}dx\,dy]\,,
\end{split}
\end{equation}
\emph{i.e.} $\gamma_{ab}=\lim_{u\rightarrow0}u^2 h_{ab}$.  As a double check, one can easily show that the following Ward identities hold\,\footnote{In general, there will be corrections from the Maxwell and axion fields to the Ward identities. In our present ansatz, those additional terms vanish and therefore are not presented to keep the presentation as simple as possible.}
\begin{equation}\label{wald}
 \gamma_{ab} T^{ab} =0,\quad \nabla_a {T^{a}}_b=0,\quad \nabla_a J^a=0\,,
\end{equation}
Moreover, $\mathcal{S}=2\pi \Sigma(u_A,t)^2/s_0(t)^2$ is the entropy density  associated with the apparent horizon. While in standard black hole thermodynamics the entropy density is defined by the area for the event horizon, for non-equilibrium cases a more appropriate one is the apparent horizon, see \emph{e.g}~\cite{Rougemont:2021gjm,Engelhardt:2017aux}. 

%%%%%%%%%%%%%%

\subsection{On the definition of strain and stress}\label{s2}
Given that the boundary metric $\gamma_{ab}$ is not flat anymore, one needs to be careful about the definition of the shear stress in response of finite deformations. Since the strain tensor is defined with respect to the original reference system, with flat metric -- our laboratory -- we need to define the stress tensor on the same footing.
Our boundary geometry is anisotropic; we find convenient to use the orthonormal frame formalism, see \emph{e.g.}~\cite{Ganguly:2021pke}. The geometry is split into a fluid moving orthogonally to the homogeneous spatial hypersurface, with the timelike fluid 4-velocity $u^a$ ($\gamma_{ab}u^au^b=-1$). For convenience, we introduce the projection tensor,
\begin{equation}
\Delta_{ab}:=\gamma_{ab}+u_a u_b\,,
\end{equation}
which represents the induced metric on the spatial hypersurface. 

The shear tensor $\tau_{ab}$ is defined as the transverse traceless symmetric part of the extrinsic curvature of the spatial hypersurface, \emph{i.e.}
\begin{equation}
K_{ab}={\Delta_{a}}^{c}{\Delta_{b}}^{d}\nabla_c u_d=K_{[ab]}+\frac{1}{2}\Theta \Delta_{ab}+\tau_{ab}\,,
\end{equation}
with $\Theta:=\gamma^{ab}K_{ab}=\nabla_a u^a$ the expansion scalar. Because of the absence of vorticity, the antisymmetric part of the extrinsic curvature vanishes, $K_{[ab]}=0$.
The symmetric energy-momentum tensor for a relativistic fluid can be written as
\begin{equation}\label{hydro}
\begin{split}
T_{ab}=\mathcal{E}u_a u_b+P\Delta_{ab}+2q_{(a}u_{b)}+\pi_{ab},
\end{split}
\end{equation}
with $\mathcal{E}=T_{ab}u^a u^b, P=\Delta^{ab}T_{ab}/2, q_a=-{h_a}^b u^c T_{bc}$, respectively, the energy density, the pressure and the heat conduction vector measured by an observer comoving with fluid. $\pi_{ab}$ is the dissipative transverse part of the stress tensor with $u^a\pi_{ab}=0$ and ${\pi^a}_a=0$. This component is related to the shear tensor $\pi_{ab}$ by
\begin{equation}\label{shear}
\pi_{ab}=-2\eta\tau_{ab}\,,  
\end{equation}
where $\eta$ is the shear viscosity.\footnote{The contribution from the bulk viscosity is set to zero because of conformal symmetry. The terms coming from a finite elastic bulk modulus are also neglected since not relevant at this point.} When the background is Minkowski, one recovers the standard constitutive relations in the Landau frame~\cite{Kovtun:2012rj}.
In our present case with metric~\eqref{Fmetric}, the fluid velocity reads $u^a=\left(\frac{\partial}{\partial t}\right)^a$, from which one finds the components of the shear tensor
\begin{equation}
\begin{split}
\tau_{xx}=\tau_{yy}=\frac{1}{2}s_0^2\dot{h}_0\sinh{(h_0)},\quad \tau_{xy}=\tau_{yx}=\frac{1}{2}s_0^2\dot{h}_0\cosh{(h_0)}\,,
\end{split}
\end{equation}
together with the volume expansion rate $\Theta=2\dot{s}_0/s_0$. For the pure shear case, the volume is unchanged, \emph{i.e.} {$\Theta=0$}. 

The stress tensor~\eqref{stress0}-\eqref{stress1} has the form of~\eqref{hydro} with $\mathcal{E}=2P=-b_3$ and $q_a=0$, and its dissipative term $\pi_{ab}$ reads
\begin{equation}
\begin{split}
\pi_{xx}&=\pi_{yy}=-s_0^2\sinh{(h_0)}\sigma\,,\\
\pi_{xy}&=\pi_{yx}=-s_0^2\cosh{(h_0)}\sigma\,,
\end{split}
\end{equation}
with 
\begin{equation}
\begin{split}
\sigma=-\frac{3}{2}h_3+ \frac{3}{2}\frac{s_1^2}{s_0^2}\dot{h}_0+\frac{1}{8}\left(3\dot{h}_0^3-4\dddot{h}_0-\frac{4\dot{h}_0\dot{s}_0^2}{s_0^2}-\frac{4(\dot{h}_0\ddot{s}_0+3\ddot{h}_0\dot{s}_0)}{s_0}\right)\,.\label{def}
\end{split}
\end{equation}
Then, we can obtain the shear viscosity 
\begin{equation}\label{eta}
\eta=-\frac{1}{2}\frac{\pi_{xy}}{\tau_{xy}}=\frac{\sigma}{\dot{h}_0}\,.
\end{equation}
We stress that the above definition on the shear viscosity is independent of the coordinates system. In contrast, each component of the stress tensor is observer dependent. Our goal is to obtain the stress in terms of an observer that experiences a flat background metric -- the laboratory. An observer, by definition, is described by a tetrad which includes a timelike unit vector $u^a$ and a set of spacelike unite vectors, defined on a spacetime manifold. In our present case, the tetrad (\emph{i.e.} local rest frame) is given by
\begin{equation}\label{observer}
\begin{split}
u^a=&\left(\frac{\partial}{\partial t}\right)^a \,,\\
v_1^a=&\frac{\cosh\left(h_0/2\right)}{s_0}\left(\frac{\partial}{\partial x}\right)^a-\frac{\sinh\left(h_0/2\right)}{s_0}\left(\frac{\partial}{\partial y}\right)^a \,,\\
v_2^a=&\frac{\sinh\left(h_0/2\right)}{s_0}\left(\frac{\partial}{\partial x}\right)^a-\frac{\cosh\left(h_0/2\right)}{s_0}\left(\frac{\partial}{\partial y}\right)^a \,,
\end{split}
\end{equation}
such that the metric tensor reads
\begin{equation}\label{gammaab}
\gamma_{ab}=-u_a u_b+v_{1a} v_{1b}+v_{2a} v_{2b}\,.   
\end{equation}
Then, one finds the non-vanishing components of $\pi_{ab}$ and $\tau_{ab}$ to be
\begin{equation}\label{right}
\pi_{12}=\pi_{ab}v_1^a v_2^b=\pi_{21}=\sigma,\quad \tau_{12}=\tau_{ab}v_1^a v_2^b=\tau_{21}=-\frac{1}{2}\dot{h}_{0}\,.
\end{equation}
Meanwhile, the pressure obtained from $P=T_{ab}v_1^a v_1^b=T_{ab}v_2^a v_2^b$ is the same as the one from~\eqref{hydro}.

The physical meaning of~\eqref{eta} is now manifest: $\sigma$ is the shear stress and $h_0\equiv \gamma$ is the shear deformation, such that one recovers the standard Newton's Law of viscosity
\begin{equation}
\underbrace{\sigma}_{\text{shear stress}}=\underbrace{\eta}_{\text{shear viscosity}}\times \underbrace{\dot{\gamma}}_{\text{shear rate}}\,.\label{eq:vis_one_order}
\end{equation}
$\gamma$ describes the mechanical deformations of the viscoelastic medium, \emph{i.e.}
\begin{equation}
d s_{\partial}^2=-d t^2+s_0(t)^2[\cosh{(\gamma(t))}(dx^2+dy^2)+2\sinh{(\gamma(t))}dxdy]\,,
\end{equation}
as material particles flow along $u^a$. Another parameter $s_0$ serves as a nonlinear version of the pure bulk deformation which will not considered in this work.

To summarize, the correct definitions for the different physical observables used in the main text are
\begin{equation}
\begin{split}
   & \gamma(t)\equiv h_0(t),\quad \Theta(t)=\frac{2\dot{s}_0(t)}{s_0(t)},\quad \rho=-\frac{\phi_1}{2} \,, \quad \mathcal{E}=2P=-b_3\,,\quad q_a=0\,, \\
   & \sigma(t)\equiv -\frac{3}{2}h_3+ \frac{3}{2}\frac{s_1^2}{s_0^2}\dot{h}_0+\frac{1}{8}\left(3\dot{h}_0^3-4\dddot{h}_0-\frac{4\dot{h}_0\dot{s}_0^2}{s_0^2}-\frac{4(\dot{h}_0\ddot{s}_0+3\ddot{h}_0\dot{s}_0)}{s_0}\right)\,.
 \end{split}
\end{equation}
Before concluding, we notice that our system displays the following scaling symmetry
\begin{equation}
(u,t,x,y)\rightarrow\lambda(u,t,x,y),\quad B\rightarrow\frac{1}{\lambda^2}B,\quad (\Sigma, \Phi)\rightarrow \frac{1}{\lambda}(\Sigma, \Phi)\,,
\end{equation}
which translates in terms of physical boundary quantities into
\begin{equation}\label{scaling}
\gamma\rightarrow\gamma,\; (\dot{\gamma},\alpha)\rightarrow\frac{1}{\lambda}(\dot{\gamma},\alpha), \; (\eta,\mathcal{S}, \rho)\rightarrow\frac{1}{\lambda^2}(\eta,\mathcal{S}, \rho),\; (\mathcal{E}, \sigma)\rightarrow\frac{1}{\lambda^3}(\mathcal{E}, \sigma)\,.
\end{equation}
This scaling symmetry is exactly what forces us to present all our results in dimensionless ratios.\\

\begin{figure}[ht]
		\begin{center}
			\includegraphics[width=0.32\textwidth]{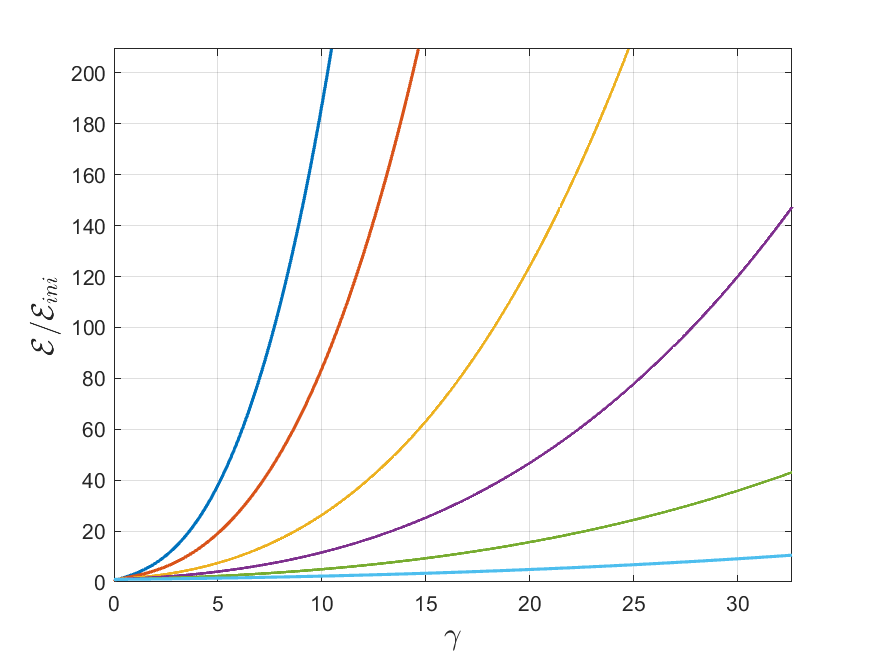}
			\includegraphics[width=0.32\textwidth]{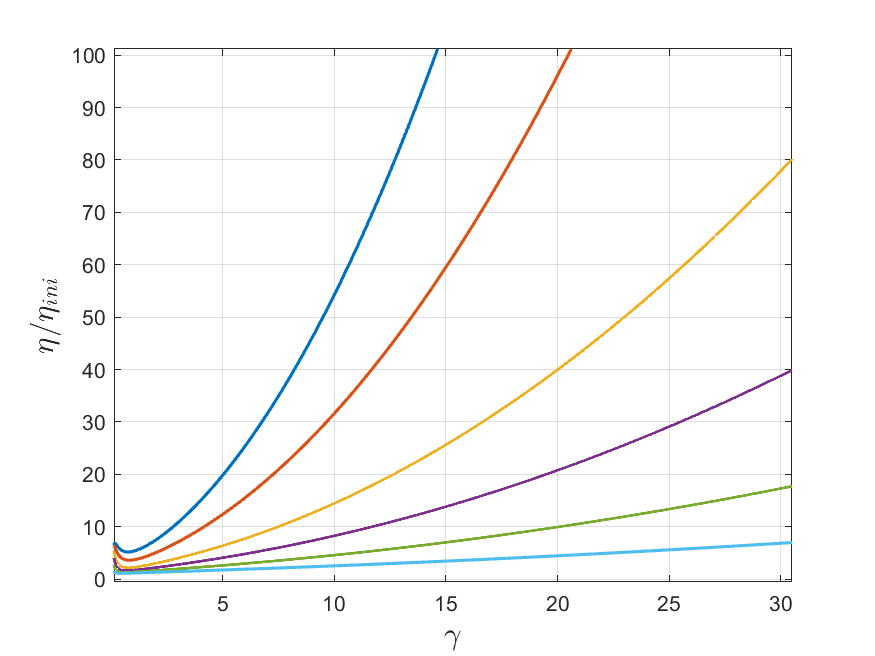}
			\includegraphics[width=0.32\textwidth]{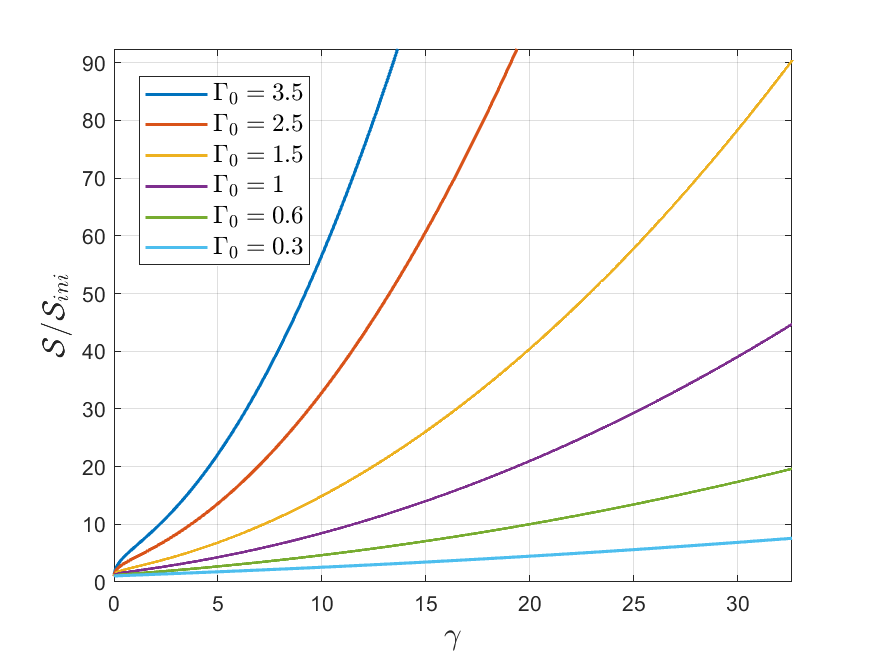}
			\caption{The far-from-equilibrium dynamics in the finite charge density EM model \eqref{model1}. The energy density $\mathcal{E}$ (\textbf{left}), the apparent viscosity $\eta$ (\textbf{center}) and the entropy density $\mathcal{S}$ (\textbf{right}) normalized by their initial near-equilibrium values as a function of the dimensionless shear strain $\gamma=\gamma_0 t$. Different colors correspond to different constant shear rates $\Gamma_0\equiv \gamma_0/\sqrt{\rho}$ from slow (light blue) to fast (dark blue).}
			\label{fig:1old}
		\end{center}
\end{figure}

Finally, let us emphasize that one could equivalently introduce a non-trivial strain into the dual field theory using the profile of the scalars $\psi^I$. In particular, a shear strain could be simply obtained by allowing for a background of the type $\psi^1 \sim y$ as done in \cite{Alberte:2018doe,Baggioli:2020qdg,Pan:2021cux}. For practical reasons, we found more convenient to introduce the strain via a metric deformation. A concrete example of the charged relativistic fluid
with constant shear rate, $\gamma=\gamma_0\, t$, is presented in Fig.~\ref{fig:1old}. The time dependent dynamics strongly depends on the dimensionless shear rate $\Gamma_0=\gamma_0/\sqrt{\rho}$ with $\rho$ the charge density of the dual field theory. For small $\Gamma_0$ the time evolution is slow and all quantities slowly grow in time. Whenever the gradients are large, all quantities rapidly increase and deviate from their initial values.

\subsection{Energy dissipation and entropy production}\label{s3}
It is interesting to see that the constraints~\eqref{eq:constraint} can be rewritten as
\begin{align}
\dot{\mathcal{E}}(t)+\frac{3\dot{s}_0(t)}{s_0(t)}\mathcal{E}(t)=\dot{\gamma}(t)\sigma(t)\,,\qquad 
\dot{\rho}(t) s_0+2\rho\dot{s}_0=0\,,
\end{align}
from which we obtain
 \begin{equation}\label{eomE}
 \begin{split}
\mathcal{E}(t)&=\frac{s_0(t_\mathrm{ini})^3}{s_0(t)^3}\left[\mathcal{E}_i+\frac{1}{s_0(t_\mathrm{ini})^3}\int_{t_\mathrm{ini}}^t s_0(\tau)^3 \dot{\gamma}(\tau)\sigma(\tau)d\tau \right]\,,\\
\rho(t)&=\frac{\rho_\mathrm{ini}}{s_0(t)^2}\,,
\end{split}
 \end{equation}
with $\mathcal{E}_\mathrm{ini}$ ($\rho_\mathrm{ini}$) the energy (charge) density at the initial time $t_\mathrm{ini}$. 

To gain some intuition, let us first consider a pure volumetric deformation by setting $\dot{\gamma}=0$. One immediately obtains from Eq.~\eqref{eomE} that
\begin{equation}\label{law}
\mathcal{E}(t) s_0(t)^3=const,\quad \rho(t) s_0(t)^2=const\,.
\end{equation}
Eq. \eqref{law} corresponds to the conservation of the charge and energy in the comoving frame of the fluid.

\begin{figure}[ht]
		\begin{center}
		\includegraphics[width=0.55\textwidth]{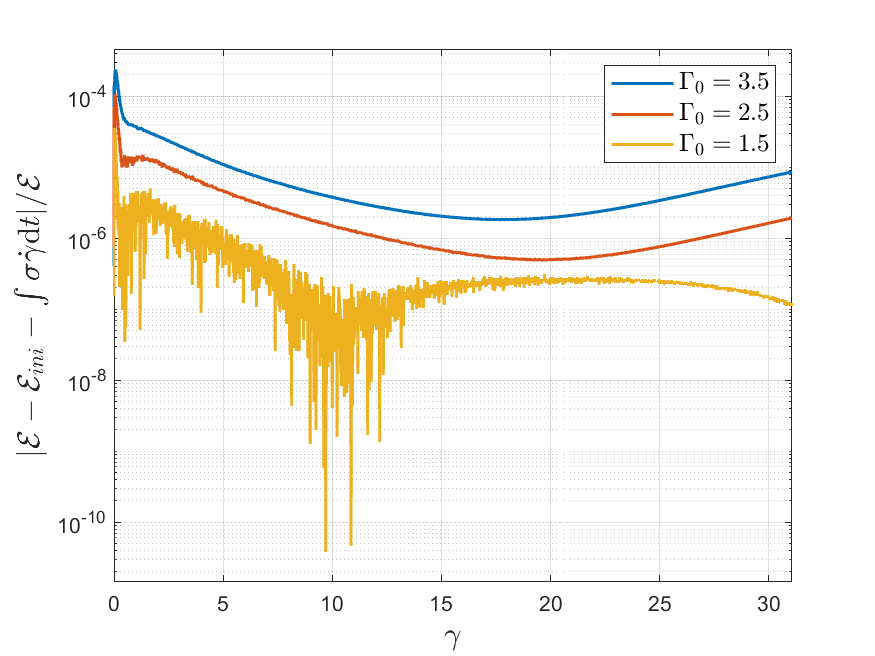}
			\caption{A numerical verification of the energy evolution equation \eqref{totest} in the EM model of \eqref{model1} and for different dimensionless shear rates.}
			\label{fig:energy_int}
		\end{center}
	\end{figure}

We are mostly interested in the pure shear case, without any volumetric deformation (\emph{i.e.} $\dot{s}_0=0$). Then, the second identity in Eq.\eqref{eomE} implies that the charge density $\rho$ is a constant. Moreover, the energy density obeys the evolution law
\begin{equation}\label{totest}
\mathcal{E}(t)=\mathcal{E}_\mathrm{ini}+\int_{t_\mathrm{ini}}^t \dot{\gamma}(\tau)\sigma(\tau) d\tau=\mathcal{E}_\mathrm{ini}+\int_{t_\mathrm{ini}}^t \eta(\tau)\dot{\gamma}(\tau)^2 d\tau\,,
\end{equation}
reported in the main text. The validity of this analytical relation is numerically verified in Fig.~\ref{fig:energy_int} in a concrete case. Moreover, the validity of \eqref{totest} can also be used as a check of the numerics.
\\
It is now manifest that the increase of the energy density is from the dissipation induced by a non-vanishing shear viscosity.

With our conventions, the entropy current density is given by $s^a=\mathcal{S} u^a$. We can then compute the entropy production using:
\begin{equation}\label{entropy}
\nabla_a s^a=\frac{2\pi}{s_0^2}\frac{\partial}{\partial t}\Sigma(u_A,t)^2\,.
\end{equation}
Notice that the requirement of the non-negativity of entropy production is equivalent to the monotonic increase of the area of the apparent horizon $\Sigma(u_A,t)^2$.\\

\begin{figure}
	   \centering
	    \includegraphics[width=0.45\textwidth]{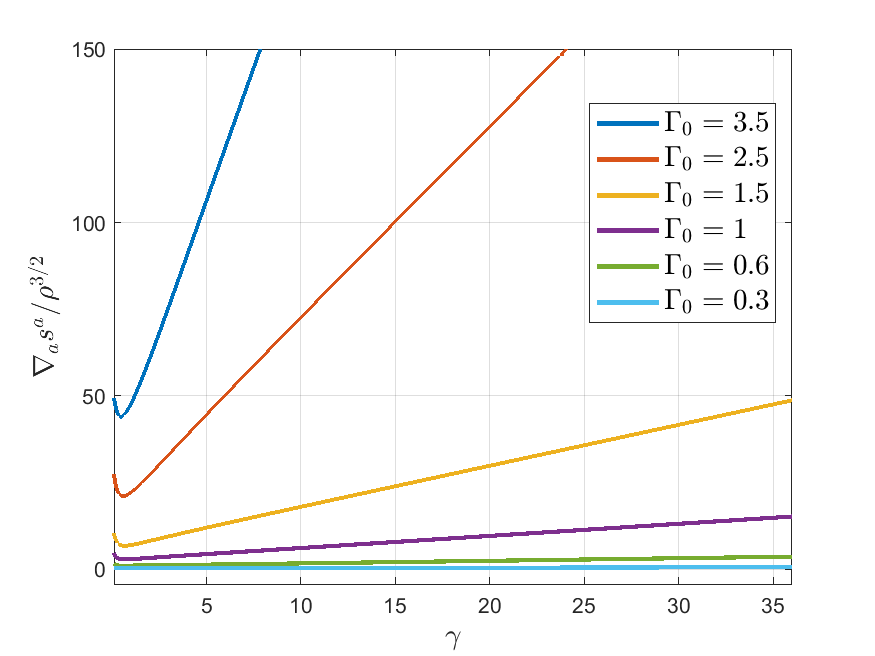}
	     \includegraphics[width=0.45\textwidth]{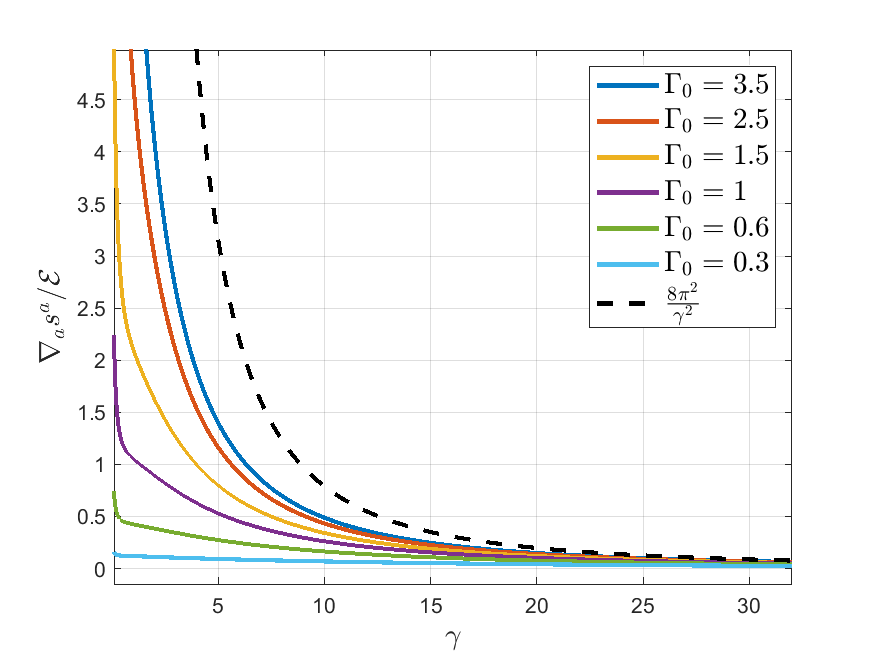}
	    \caption{\textbf{Left: } Entropy production as a function of the dimensionless shear deformation $\gamma$ for different rates $\Gamma_0$. \textbf{Right: } The entropy production to energy ratio $\nabla_a s^a/\mathcal{E}$ as a function of $\gamma$. The large $\gamma$ region follows a power-law scaling $\sim 1/\gamma^2$.}
	    \label{fig:entropy}
\end{figure}

In the left panel of Fig.~\ref{fig:entropy}, we show the entropy production as a function of the dimensionless shear deformation $\gamma$ for different shear rates $\Gamma_0$. The entropy production displays a mild peak around a small value of $\gamma$. This value grows very mildly with the shear rate $\Gamma_0$ following the same trend of the dip in the $\eta/s$ ratio in Fig.~\ref{fig:1} and of the maximum in the shear rate to energy ratio in Fig.~\ref{fig:2}. After that initial stage, the entropy production reaches a large $\gamma$ regime where it is linear in $\gamma$ with the slope increasing with $\Gamma_0$. More precisely, we find from our numerics that
\begin{equation}
    \nabla_a s^a \,=\,\frac{\gamma_0^3}{\pi}\,\gamma\,=\,\frac{\gamma_0^4}{\pi}\,t\,\qquad \text{for}\qquad \gamma \gg 1\,.
\end{equation}
In the right panel of Fig.~\ref{fig:entropy}, we normalize the entropy production rate in terms of the energy density of the system. There, we notice that for large $\gamma$ the entropy production normalized in this way goes to zero as $1/\gamma^2$, confirming that energy is the dominant scale at large $\gamma$, when the universal attractor solution in Eq.\eqref{attra} is reached.

\subsection{Different holographic models and different shear rates}\label{s4}
In order to confirm the universal character of the observations presented in the main text, in this Section we extend the computations to the second holographic model defined in Eq. \eqref{model2}. Moreover, for both the holographic models \eqref{model1} and \eqref{model2}, we also consider different shear deformations with time dependent shear rates.

\begin{figure}[ht]
		\begin{center}
		\includegraphics[width=0.45\textwidth]{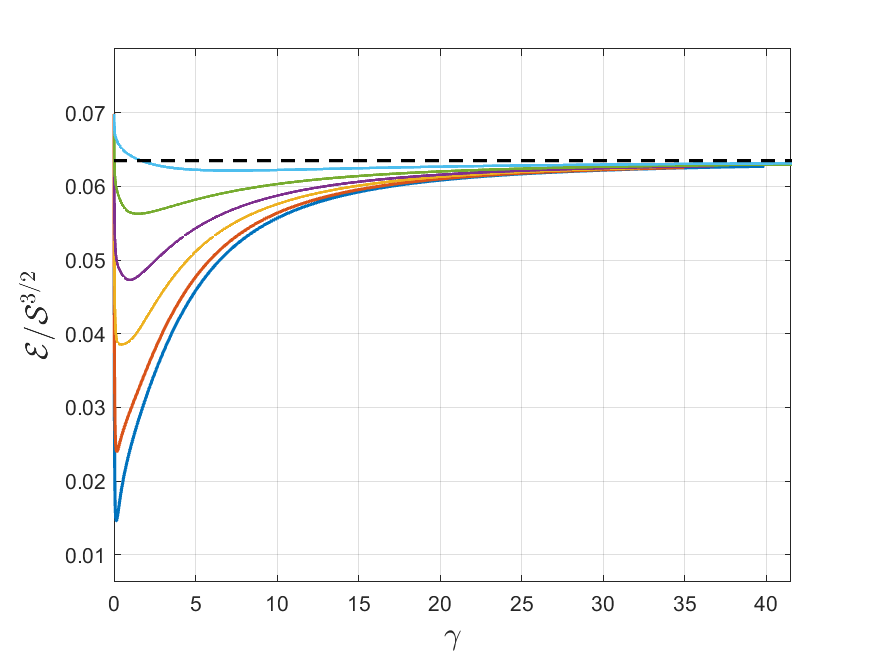}
		\includegraphics[width=0.45\textwidth]{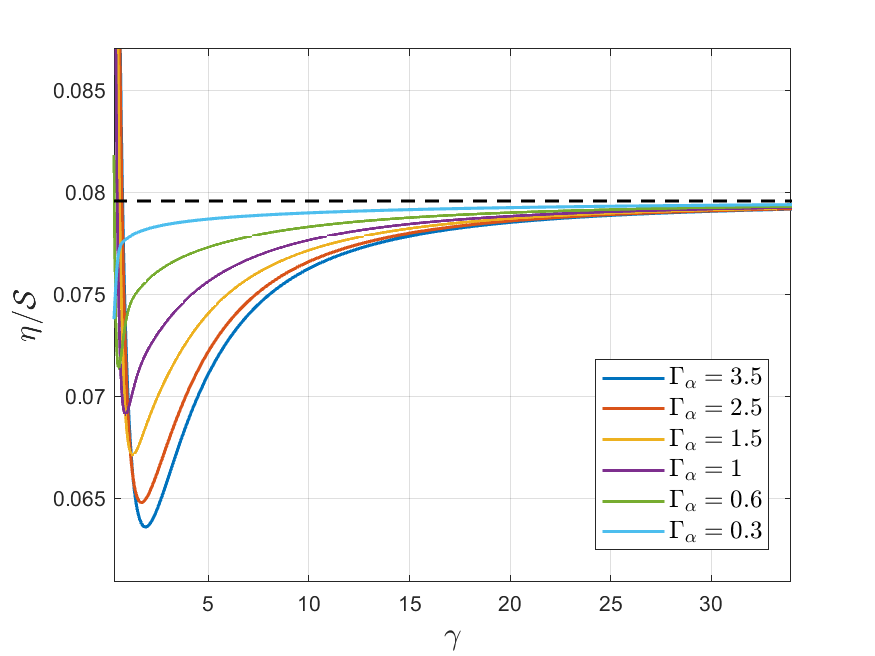}
			\caption{The universal behavior in the second holographic model of Eq.\eqref{model2} with constant shear rate $\gamma=\gamma_0 t$ and $\Gamma_\alpha\equiv \gamma_0/\alpha$. \textbf{Left: } $\mathcal{E}/\mathcal{S}^{3/2}$ as a function of $\gamma(t)$, where the dash line is $1/(2\pi)^{3/2}$. \textbf{Right: } $\eta/\mathcal{S}$ as a function of $\gamma(t)$, where the dash line is $1/4\pi$.}
			\label{fig:s1}
		\end{center}
\end{figure}

We start by considering the model of Eq.\eqref{model2} with a constant shear rate $\gamma(t)=\gamma_0 t$. We present our results in terms of the dimensionless rate:
\begin{equation}
    \Gamma_\alpha \equiv \gamma_0/\alpha\,,
\end{equation}
where $\alpha$ is the scale that parametrizes the bulk modulus of the fluid. As evident from the curves in Fig.~\ref{fig:s1}, we confirm that the same late time behavior is recovered also in this model, and both $\eta/\mathcal{S}$ and $\mathcal{E}/\mathcal{S}^{3/2}$ tend at late time, and for large gradients, to the same universal constants reported in Eq.\eqref{attra}.

\begin{figure}[ht]
		\begin{center}
		\includegraphics[width=0.32\textwidth]{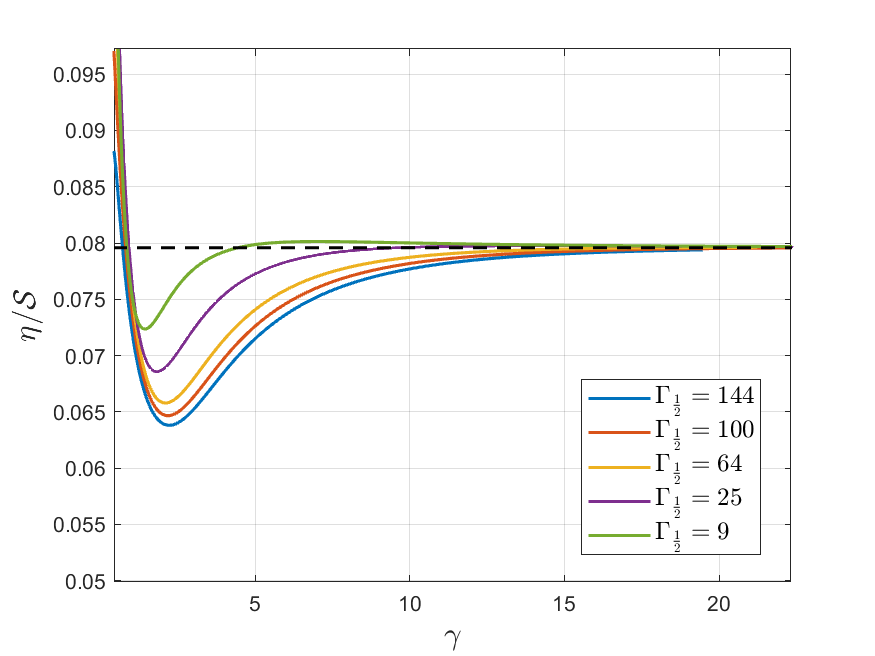}
		\includegraphics[width=0.32\textwidth]{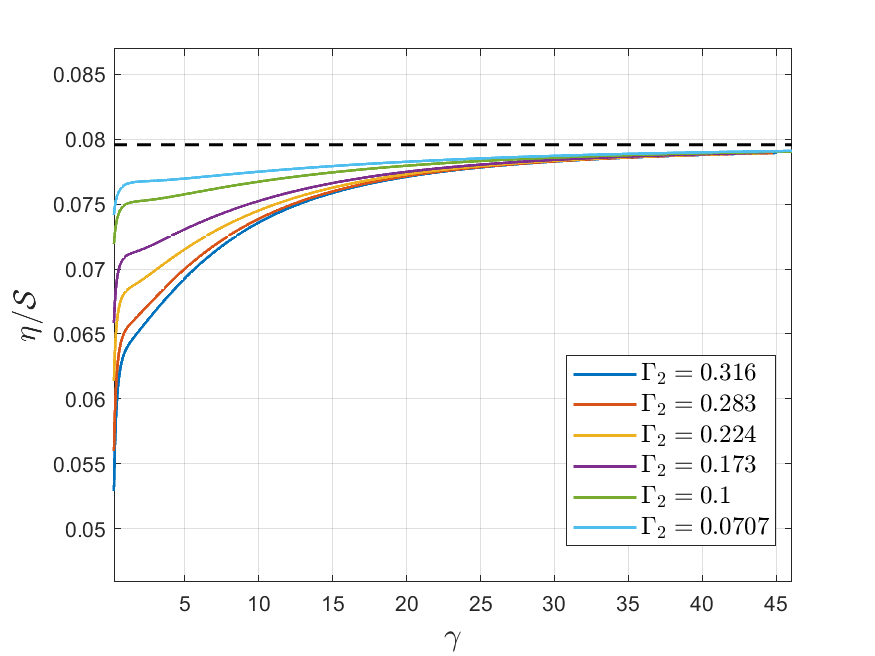}
		\includegraphics[width=0.32\textwidth]{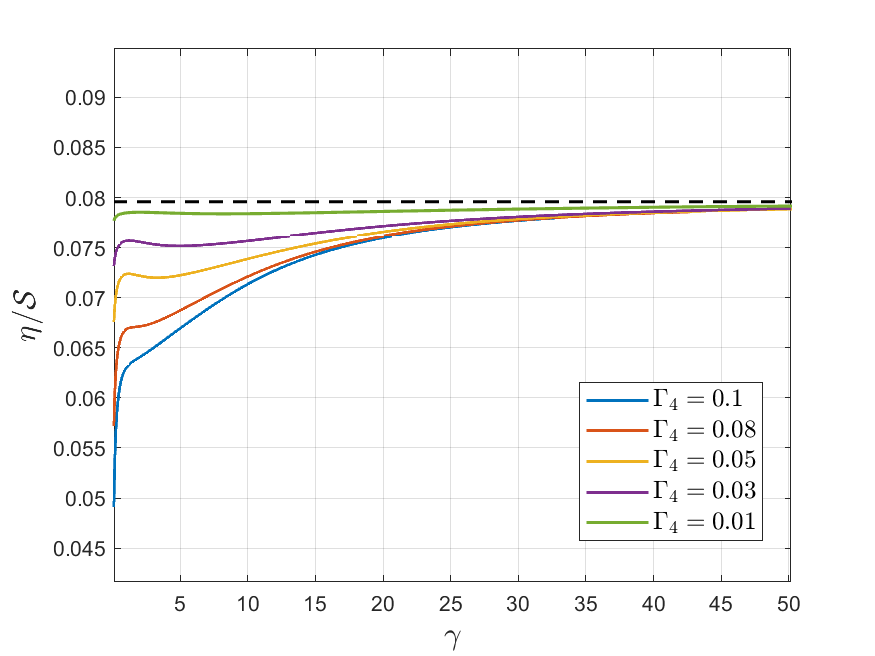}
		
		\includegraphics[width=0.32\textwidth]{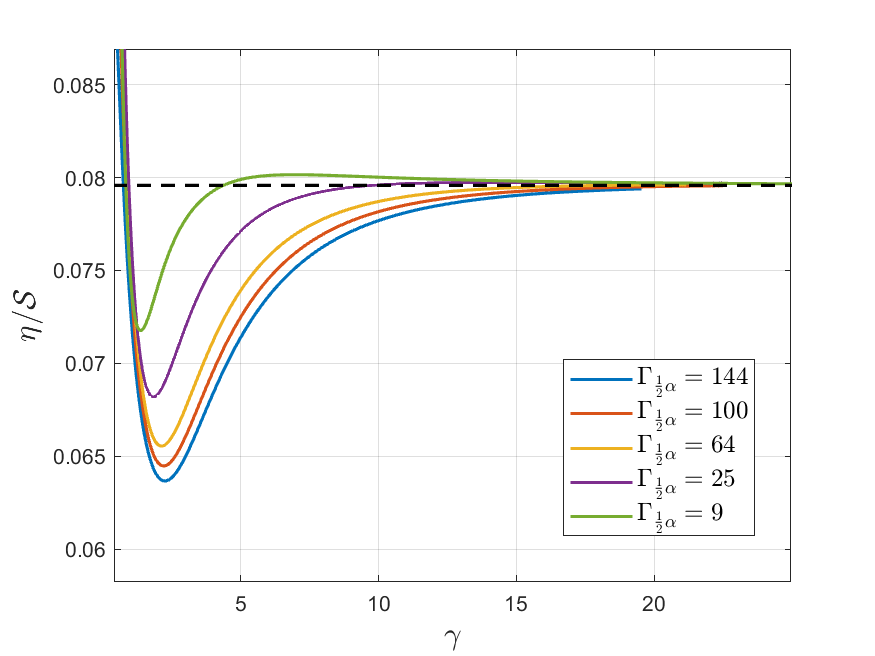}
		\includegraphics[width=0.32\textwidth]{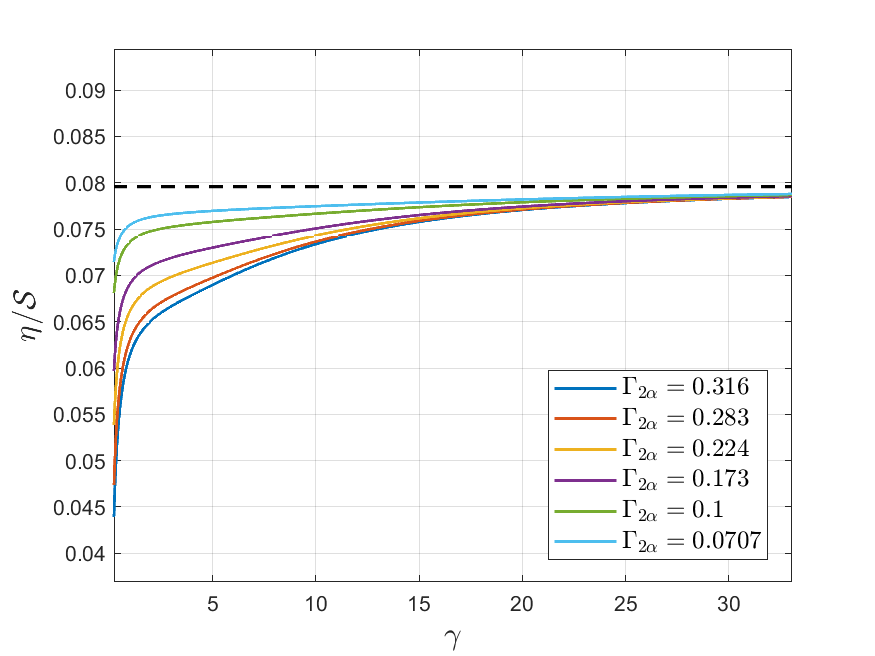}
		\includegraphics[width=0.32\textwidth]{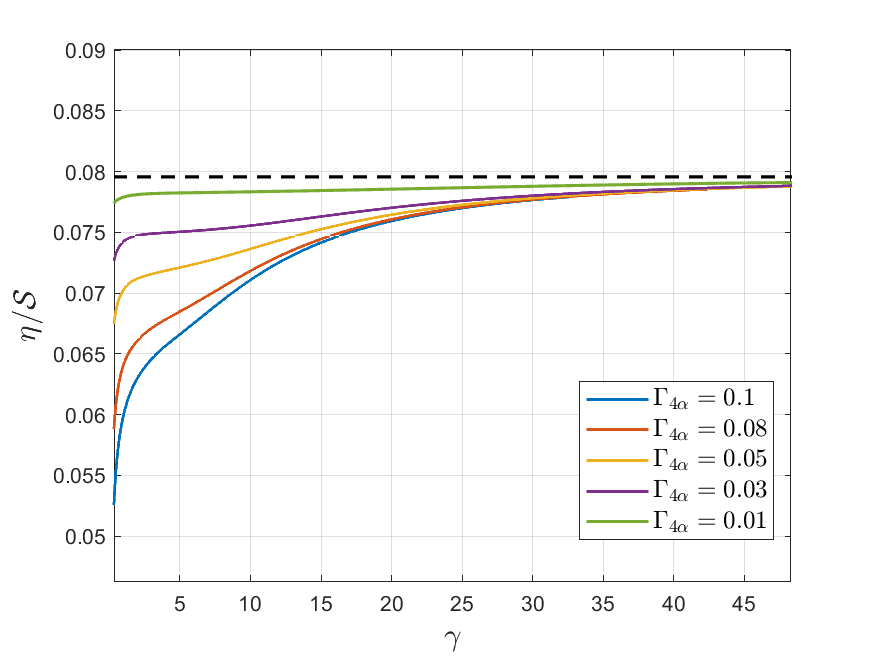}
			\caption{The top panels refer to the charged model of Eq.\eqref{model1}, while the bottom panels to the neutral model of Eq.\eqref{model2}. All figures concern the dimensionless ratio $\eta/\mathcal{S}$. We have defined $\Gamma_{1/2}\equiv \gamma_{1/2}/\sqrt{\rho}$, $\Gamma_{2}\equiv \gamma_{2}/\sqrt{\rho}$, $\Gamma_{4}\equiv \gamma_{4}/\sqrt{\rho}$, $\Gamma_{1/2 \alpha}\equiv \gamma_{1/2}/\alpha$, $\Gamma_{1/2\alpha}\equiv \gamma_{1/2}/\alpha$, $\Gamma_{4\alpha}\equiv \gamma_{4}/\alpha$. \textbf{Top Left: } Strain function \eqref{strain1}. \textbf{Top Center: } Strain function \eqref{strain2}. \textbf{Top Right: } Strain function \eqref{strain3}. \textbf{Bottom Left: } Strain function \eqref{strain1}. \textbf{Bottom Center: } Strain function \eqref{strain2}. \textbf{Bottom Right: } Strain function \eqref{strain3}.}
			\label{fig:s2}
		\end{center}
	\end{figure}

To provide even more evidence for this universality, we consider three different types of shear strain deformations
\begin{align}
&{\gamma}(t)=\sqrt{\gamma_{1/2}\,t} \label{strain1}\,,\\
&{\gamma}(t)=(\gamma_{2}\, t)^2\label{strain2}\,,\\
&{\gamma}(t)=(\gamma_4 t)^4 \label{strain3}\,,
\end{align}
which do not involve a constant shear rate $\dot{\gamma}$. Here $\gamma_{1/2}, \gamma_{2}$ and $\gamma_{4}$ are all constants. Again, we find that the same universality emerges for arbitrary shapes of the strain deformations. The results for the viscosity-to-entropy density ratio are presented in Fig.~\ref{fig:s2}. Similar results are found for the dimensionless ratio $\mathcal{E}/\mathcal{S}^{3/2}$.

    %%%%%%%%%%%%%%%%%%%%%%%%%%%%%%%%%%%%
\subsection{Non-conformal case}\label{snonconf}
So far, we have considered holographic setups which enjoy conformal invariant and whose stress-tensor is traceless, ${T_a}^a=0$. In order to break conformal invariance in the boundary field theory, a simple way is to turn on a relevant deformation via a scalar operator. In the bulk side, one introduces a real scalar field with an explicit source at the AdS boundary.

We consider the four-dimensional bulk action~\cite{Li:2020spf} :
\begin{equation}\label{model3}
S\,=\frac{1}{2\kappa_N^2}\,\int d^4x \sqrt{-g}
\left[\mathcal{R}+6-\frac{1}{2}\nabla_a\psi\nabla^a\psi+\frac{4(\sinh{\delta \psi})^2}{\delta^2}\right]\,,
\end{equation}
with $\delta$ a constant and $\kappa_N=1$. Under the ansatz~\eqref{appansatz} together with $\psi=\psi(u,t)$, the bulk equations of motion are given by
\begin{equation}
\begin{split}
%\begin{eqnarray}
&\Sigma{}^{''} +\frac{2}{u}\Sigma{}^{'}+\frac{{H{}^{'}}^2 }{4}\Sigma+\frac{\Sigma\psi'^2}{4}=0\,,\\
&(d_+\Sigma){}^{'} +\frac{\Sigma{}^{'} }{\Sigma } {d_+\Sigma}=-\frac{3\Sigma}{2 u^2}-\frac{\left[-1+\cosh{(\delta\phi)}\right]\Sigma}{2\delta^2u^2}\,,\\
&(d_+H){}^{'} +\frac{ \Sigma^{'} }{\Sigma}d_+H=-\frac{H{}^{'} {d_+\Sigma} }{\Sigma }\,,\\
&B{}^{''}+\frac{2 }{u}B{}^{'}= \frac{H{}^{'}{d_+H}} {u^2}-\frac{4\Sigma{}^{'}{d_+\Sigma}}{u^2 \Sigma^2}+\frac{d_+\psi \psi'}{u^2}\,,\\
&4{d_+^2\Sigma} +2 u^2 B{}^{'}d_+\Sigma+ {\left(d_+H^2+d_+\psi^2\right)} \Sigma+B\left[4(d_+\Sigma)'u^2+\frac{2(-1+3\delta^2+\cosh{\delta\psi})\sigma}{\delta^2}+\frac{4d_+\Sigma u^2\Sigma'}{\Sigma}\right] =0\,,\\
&(d_+\psi)'=\frac{\sinh{\delta\psi}}{u^2\delta}+\frac{d_+\Sigma\psi'+d_+\psi\Sigma'}{\Sigma}.
%\end{eqnarray}
\end{split}
\end{equation}
The asymptotics at the AdS boundary are found to be
\begin{equation}
\label{aas}
\begin{split}
&B=\frac{1}{u^2}\left[1+\frac{2(s_1-\dot{s}_0)}{s_0}u+\left( \frac{s_1^2}{s_0^2}-\frac{2\dot{s}_1}{s_0}-\frac{3\dot{h}_0^2}{4} -\frac{\psi_1^2}{4}\right)u^2+b_3 u^3+\mathcal{O}(u^4)\right],\\
&\Sigma=\frac{1}{u}\left[s_0+s_1 u-\frac{s_0\left(\dot{h}_0^2+\psi_1^2\right)}{8}u^2+\frac{\left(3s_1 \dot{h}_0^2-\psi_1^2s_1-4 \psi_1 \psi_2s_0\right)}{24} u^3+\mathcal{O}(u^3)\right],\\
&H=h_0+\dot{h}_0 u-\frac{s_1 \dot{h}_0}{s_0}u^2+h_3 u^3+\mathcal{O}(u^4)\,,\\
&\psi=\psi_1 u+\psi_2 u^2+\mathcal{O}(u^3)\,,
\end{split}
\end{equation}
together with the constraint:
\begin{equation}\label{eq:constraintpsi}
\begin{split}
\dot{b}_3+\frac{3 b_3 \dot{s}_0}{s_0}-\dot{h}_0^2 \left(\frac{\dot{s}_0^2-3 s_1^2}{2 s_0^2}+\frac{\ddot{s}_0}{2 s_0}-\frac{\psi_1^2}{8}\right)-\frac{3 \dot{h}_0 \ddot{h}_0 \dot{s}_0}{2 s_0}
+\frac{3}{8} \dot{h}_0^4-\frac{3}{2} h_3 \dot{h}_0-\frac{1}{2} \dddot{h}_0 \dot{h}_0\\
+\psi_1\left(\frac{s_1\dot{\psi}_1}{2s_0}+\frac{\dot{\psi}_2}{2}+\frac{\psi_2\dot{s}_0}{s_0}-\frac{3\dot{\psi}_1\dot{s}_0}{2s_0}-\frac{\ddot{\psi}_1}{2}\right)-\psi_1^2\left(\frac{s_1\dot{s}_0}{2s_0^2}-\frac{\dot{s}_0^2}{2s_0^2}+\frac{\dot{s}_1}{2s_0}-\frac{\ddot{s}_0}{2s_0}\right)=0\,.
\end{split}
\end{equation}
The boundary stress tensor $T_{ab}$ can be obtained by computing~\cite{Li:2020spf}
\begin{equation}
T_{ab}=\lim_{u\rightarrow 0}-\frac{2}{u\sqrt{-h}}\frac{\delta S_{\text{ren}}}{\delta h^{ab}}=\frac{1}{2\kappa_N^2}\lim_{u\rightarrow 0}\frac{2}{u}\,\left(\mathcal{K} \,h_{ab}-\mathcal{K}_{ab}-2h_{ab}+G^h_{ab}-\frac{1}{4}\psi^2h_{ab}\right)\,,
\end{equation}
and the non-vanishing components are found to be
\begin{align}
\mathcal{E} \equiv& \,T_{tt} ={ -b_3}-\frac{1}{2}\psi_1 \psi_2-\frac{\psi_1^2 s_1}{2 s_0}+\frac{\psi_1\dot{\psi}_1}{2}+\frac{\psi_1^2\dot{s}_0}{2s_0}\,,\\
T_{xx}=&\,T_{yy}=-\frac{1}{2}b_3\cosh(h_0)s_0^2\notag\\
&\qquad+\frac{1}{8}\sinh(h_0)\left(12h_3 s_0^2-\psi_1^2 s_0^2\dot{h}_0-12s_1^2\dot{h}_0-3{s_0}^2\dot{h}_0^3+4\dot{h}_0\dot{s}_0^2+12s_0\dot{s}_0\ddot{h}_0+4s_0\dot{h}_0\ddot{s}_0+4s_0^2\dddot{h}_0\right)\\
T_{xy}=&\,T_{yx}=-\frac{1}{2}b_3\sinh(h_0)s_0^2\notag\\
&+\cosh(h_0)\left(12h_3 s_0^2-\psi_1^2 s_0^2\dot{h}_0-12s_1^2\dot{h}_0-3s_0^2\dot{h}_0^3+4\dot{h}_0\dot{s}_0^2+12s_0\dot{s}_0\ddot{h}_0+4s_0\dot{h}_0\ddot{s}_0+4s_0^2\dddot{h}_0\right)\,.
\end{align}
Moreover, the expectation value of the scalar operator reads~\cite{Li:2020spf}
\begin{equation}\label{waldpsi}
\langle\mathcal{O}_\psi\rangle=\frac{\psi_2}{2}-\frac{\dot{\psi}_1}{2}+\frac{\psi_1 s_1-\psi_1\dot{s}_0}{2s_0}\,
\end{equation}
while its source is given by the leading coefficient in the asymptotic expansion Eq.\eqref{aas}, $\psi_1$.
It is straightforward to check that the above stress tensor satisfies the known Ward identities:
\begin{equation}
\nabla_a {T^{a}}_b=\left<\mathcal{O}_\psi\right>\nabla_b \psi_1,\quad {T^{a}}_{a}=\left<\mathcal{O}_\psi\right> \psi_1\,.
\end{equation}
As expected, the boundary stress tensor has now a non-vanishing trace due to the presence of the source term $\psi_1$ for the bulk scalar $\psi$.

\begin{figure}[ht]
		\begin{center}
		\includegraphics[width=0.32\textwidth]{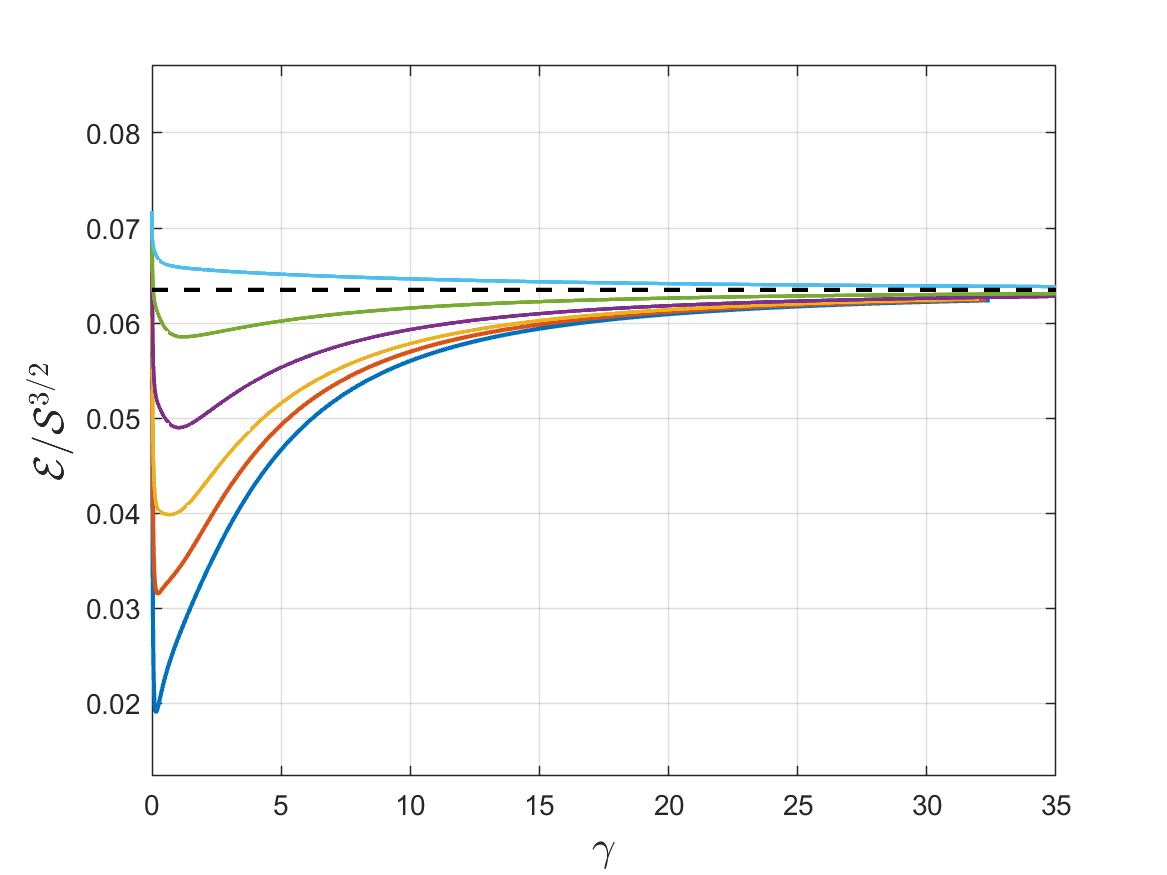}
		\includegraphics[width=0.32\textwidth]{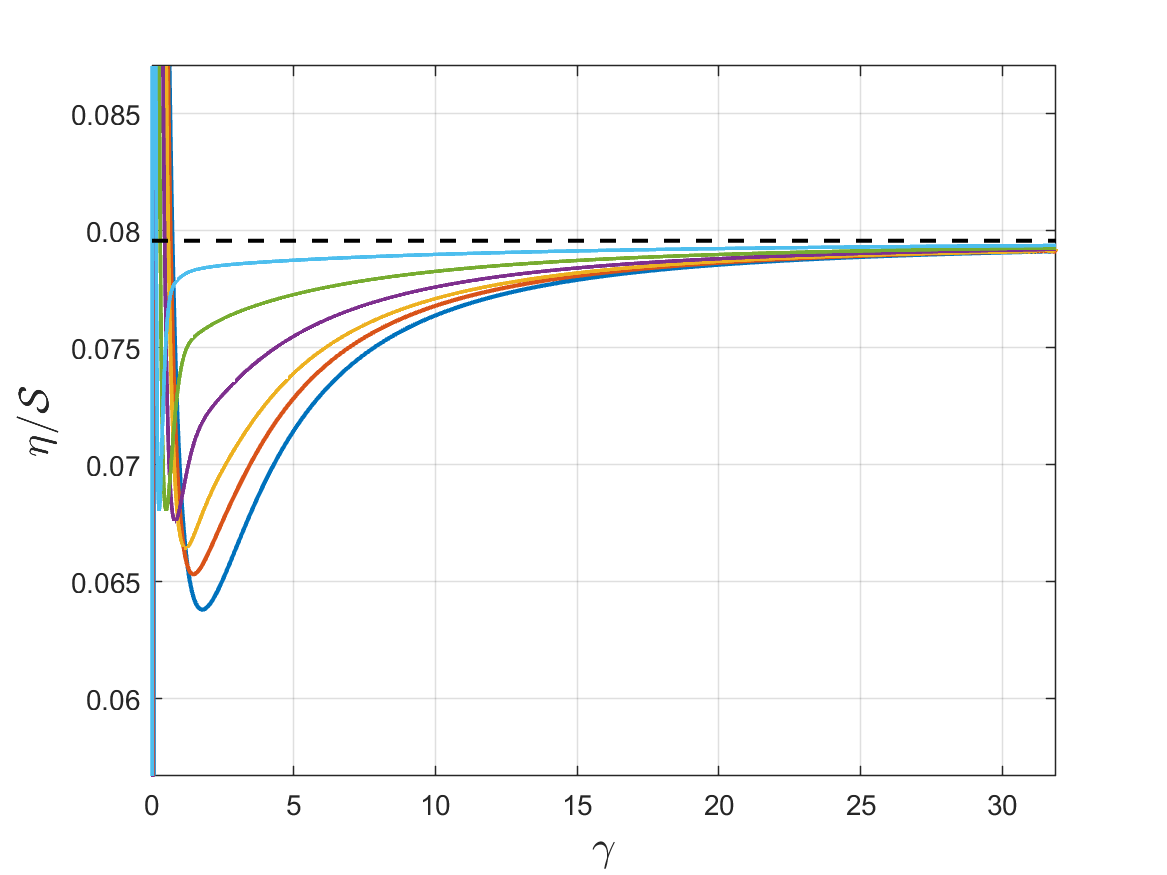}
		\includegraphics[width=0.32\textwidth]{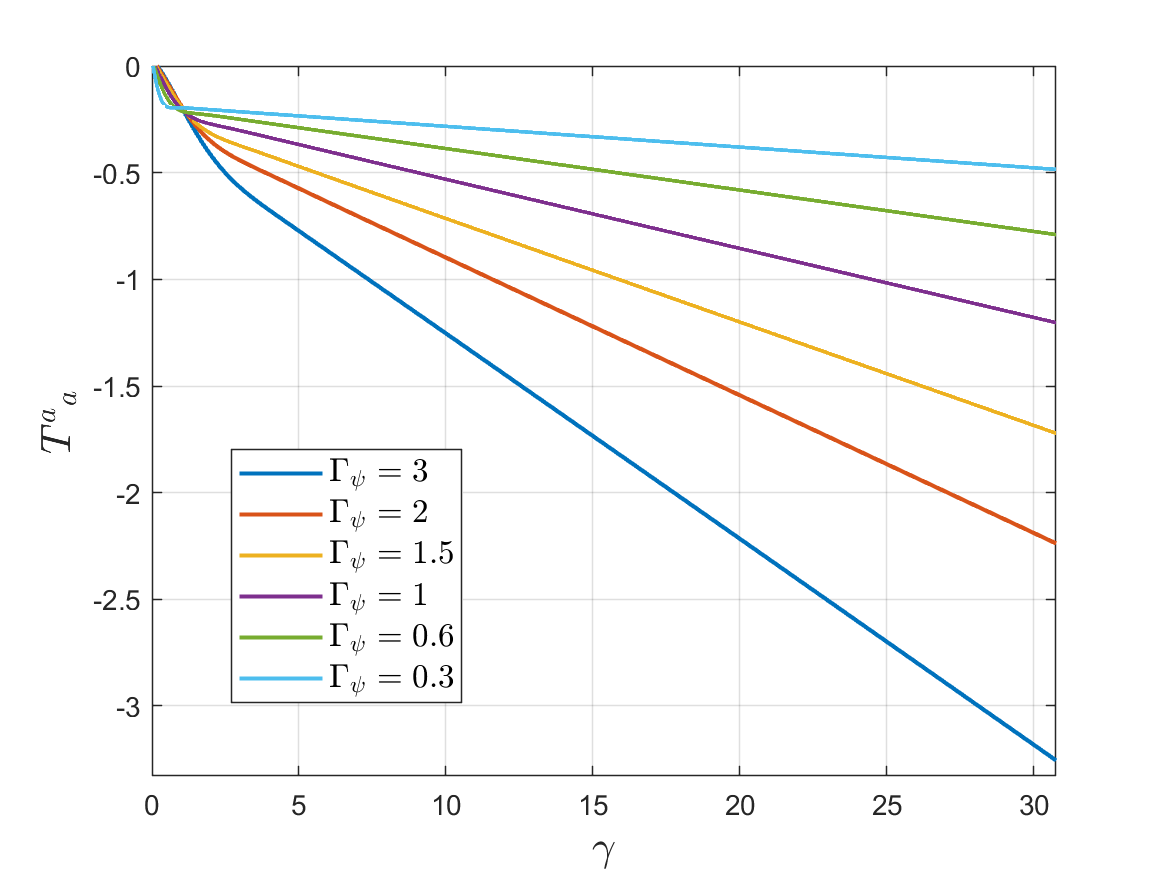}
			\caption{The non-conformal holographic model of Eq.\eqref{model3} with constant shear rate $\gamma=\gamma_0 t$. The dimensionless shear rate is indicated as $\Gamma_\psi \equiv \gamma_0/\psi_1$. \textbf{Left: } $\mathcal{E}/\mathcal{S}^{3/2}$ as a function of $\gamma(t)$, where the dash line is $1/(2\pi)^{3/2}$. \textbf{Center: } $\eta/\mathcal{S}$ as a function of $\gamma(t)$, where the dash line is $1/4\pi$. \textbf{Right: }The trace of stress tensor $T_{ab}$.}
			\label{fig:noncft}
		\end{center}
    \end{figure}

Closely following the discussion in section~\ref{s2}, we obtain the shear stress 
\begin{align}
    \sigma=-\frac{3}{2}h_3+ \frac{3}{2}\frac{s_1^2}{s_0^2}\dot{h}_0+\frac{1}{8}\psi_1^2\dot{h}_0+\frac{1}{8}\left(3\dot{h}_0^3-4\dddot{h}_0-\frac{4\dot{h}_0\dot{s}_0^2}{s_0^2}-\frac{4(\dot{h}_0\ddot{s}_0+3\ddot{h}_0\dot{s}_0)}{s_0}\right)\,.
\end{align}
The equation describing energy dissipation is also modified into
\begin{align}\label{noncft}
\dot{\mathcal{E}}+\frac{3\dot{s}_0}{s_0}\mathcal{E}=\dot{\gamma}\sigma-\frac{\left(\psi_1\dot{s}_0+s_0\dot{\psi}_1\right)\left(s_1-\dot{s}_0\right)\left<\mathcal{O}_\psi\right>}{s_0}\,.
\end{align}
Note that additional terms appear due to the relevant deformation driven by the scalar operator dual to $\psi$. Nevertheless, for the pure shear case (\emph{i.e.} $\dot{s}_0=0$) and for the deformation with a constant source (\emph{i.e.} $\dot{\psi}_1=0$), the last term in the right hand side of~\eqref{noncft} vanishes and the energy density obeys the same evolution law~\eqref{totest} reported in the main text. In the present work, we are interested in the cases with unit value for the source of the scalar field. As shown in Fig.~\ref{fig:noncft}, we find the same universality for $\eta/s$ and $\mathcal{E}/\mathcal{S}^{3/2}$, even through the conformal symmetry is broken (\emph{i.e.} the stress tensor has a non-vanishing trace). Notice also that the deviation from conformality, ${T^a}_a$, grows with time as shown in the right panel of Fig.~\ref{fig:noncft}.
%%%%%%
%%%%%%%%%%%%

\subsection{The far-from-equilibrium late time steady state}\label{sfar}
\begin{figure}[ht]
\begin{center}
\includegraphics[width=0.45\textwidth]{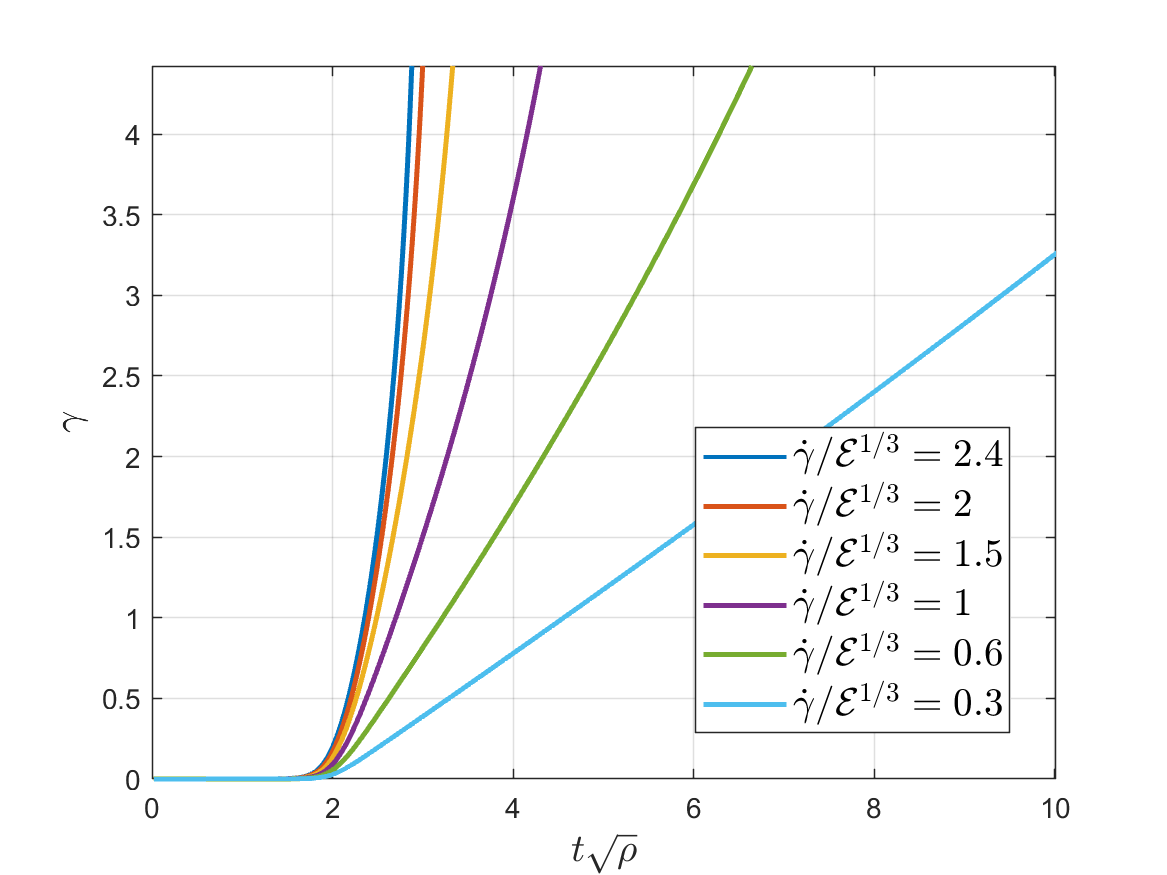}
\includegraphics[width=0.45\textwidth]{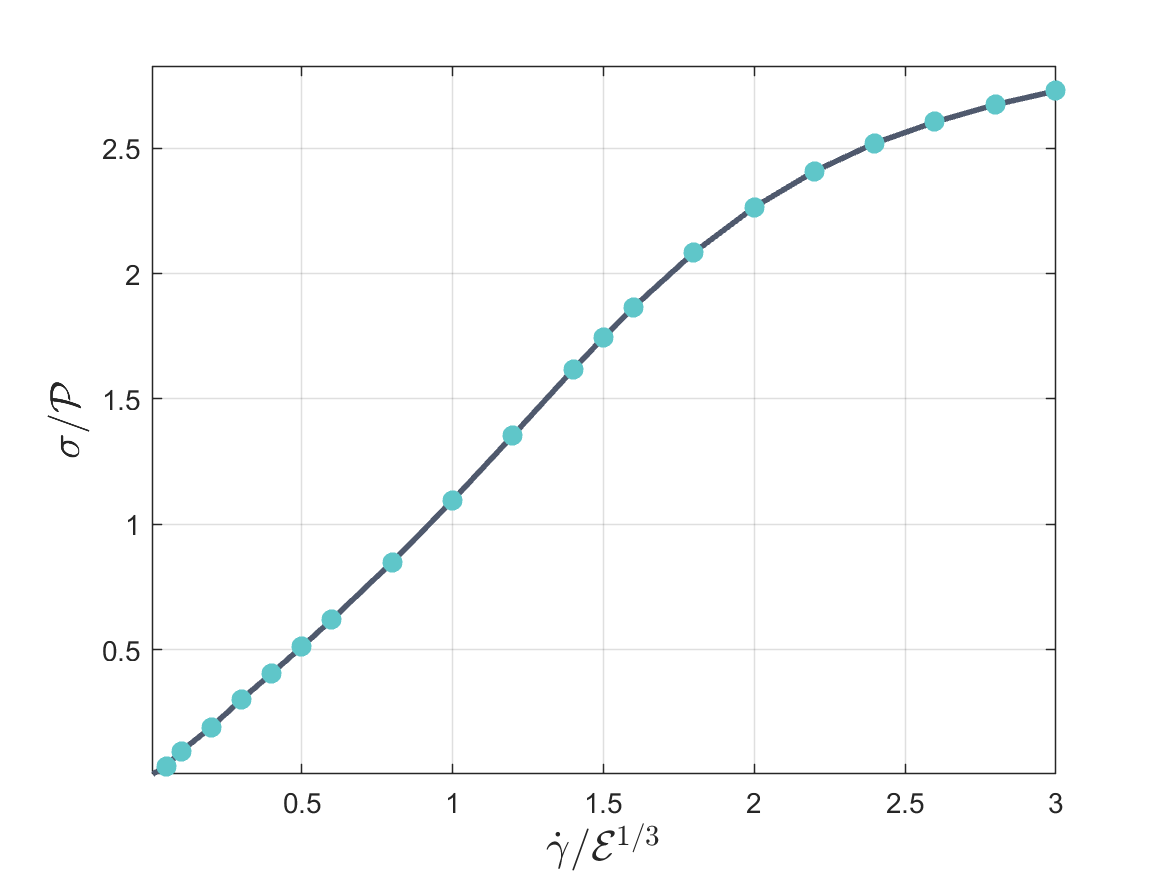}
\caption{\textbf{Left: }The shear strain $\gamma$ as a function of $t$ for different dimensionless shear rate $\dot{\gamma}/\mathcal{E}^{1/3}$ in the finite charge density EM model~\eqref{model1}. \textbf{Right: }Late time value of $\sigma/\mathcal{P}$ as a function of $\dot{\gamma}/\mathcal{E}^{1/3}$ signaling the onset of far-from-equilibrium dynamics.}
\label{fig:gammaP}
\end{center}
\end{figure}

In all above cases, the shear rate in units of the characteristic energy scale of the system becomes small at late time (see top panel of Fig.\ref{fig:2} in the main text), suggesting the presence of a late time steady state which is effectively in equilibrium. In order to have a far-from-equilibrium late time state, we consider a different time-dependent shear rate by keeping the energy-normalized shear rate fixed. In this case, the shear strain $\gamma(t)$ can only be obtained by solving the equations of motion. $\gamma$ as a function of $t$ for different $\dot{\gamma}/\mathcal{E}^{1/3}$ is shown in the left panel of Fig.~\ref{fig:gammaP}. One can see that $\gamma$ increases much faster with time compared to the constant shear rate case ($\gamma=\gamma_0 t$).

The system never reaches an effective equilibrium state where all the quantities are
time dependent but the gradients are small compared to the energy scale set by $\mathcal{E}$. To confirm the out-of-equilibrium nature of our time dynamics, we consider the pressure anisotropy $\Delta \mathcal{P}$ at late time. More precisely, 
the stress tensor of our system in the 2-dimensional spatial subspace is given by
\begin{equation}
    T_{ij}=\begin{pmatrix}
\mathcal{P} & \sigma \\
\sigma & \mathcal{P} 
\end{pmatrix}\,,
\end{equation}
with $\mathcal{P}$ and $\sigma$, respectively, the isotropic pressure and shear stress induced by the time-dependent deformation. By a simple change of reference frame which diagonalizes the above matrix, one obtains~\footnote{Compared to the the reference $(u^a, v_1^a, v_2^a)$ of~\eqref{observer}, the stress tensor appears diagonal by choosing the new reference $(u^a, \tilde{v}_1^a, \tilde{v}_2^a)$ where $\tilde{v}_1^a=(v_1^2+v_2^a)/\sqrt{2}$ and $\tilde{v}_1^a=(v_1^2-v_2^a)/\sqrt{2}$. }
\begin{equation}
   \tilde  T_{ij}=\begin{pmatrix}
P_1 & 0 \\
0 & P_2 
\end{pmatrix} \qquad \text{with}\qquad P_1=\mathcal{P}+\sigma,\,\,\quad \,P_2=\mathcal{P}-\sigma\,.
\end{equation}
In this way, we can define the normalized pressure anisotropy:
\begin{equation}\label{pp}
    \Delta \mathcal{P} \equiv \frac{P_1-P_2}{P_1+P_2}=\frac{\sigma}{\mathcal{P}}\,.
\end{equation}
As shown in the bottom panel of Fig.~\ref{fig:2} in the main text, the pressure anisotropy for a finite $\dot{\gamma}/\mathcal{E}^{1/3}$ reaches a constant and large value at late time, indicating that the system is indeed in a far from equilibrium state. 

In order to prove this further, we show the pressure anisotropy as a function of the dimensionless shear gradient $\dot{\gamma}/\mathcal{E}^{1/3}$ in the right panel of Fig.~\ref{fig:gammaP}. We find that the late time pressure anisotropy grows monotonically as the dimensionless shear gradient is increased. This indicates that, as expected, both $\dot{\gamma}/\mathcal{E}^{1/3}$ (which is now kept fixed during the simulations) and the normalized pressure anisotropy defined in Eq.\eqref{pp} are good probes for the out-of-equilibrium nature of the system with qualitatively similar behaviors. Interestingly, at large $\dot{\gamma}/\mathcal{E}^{1/3}$, the late time pressure anisotropy seems to saturate to a constant, suggesting the existence of an upper bound on the "distance from equilibrium" that can be reached in our setup. Unfortunately, due to the limitation of computing power, we are not able to confirm this upper bound.

\subsection{Numerical methods}\label{s5}
In this Section, we provide more details about the numerical techniques employed. It is convenient to use the following re-definitions
\begin{align}
    &B=\frac{1+\tilde{B}u}{u^2},\qquad \Sigma=\frac{\tilde{\Sigma}}{u},\qquad H=h_0+\tilde{H}u,\qquad d_+\Sigma=\frac{\widetilde{d_+\Sigma}}{u^2},\qquad
    d_+H=\frac{\dot{h}_0}{2}+\widetilde{d_+H}u\,,
\end{align}
such that the new functions are finite at the AdS boundary.
The new equations of motion takes the form
\begin{align}
    &4\tilde{\Sigma}''+4\tilde{\Sigma}(\tilde{H}+u\tilde{H}')^2=0\,,\label{Cons1}\\
    &u\Phi''+\frac{2u\tilde{\Sigma}'\Phi'}{\tilde{\Sigma}}=0\,,\label{Cons2}\\
    &u\widetilde{d_+\Sigma}'+(-3+\frac{u\tilde{\Sigma}'}{\tilde{\Sigma}})\widetilde{d_+\Sigma}=\frac{u^8\alpha^8}{4\tilde{\Sigma}^7}+\frac{\tilde{\Sigma}}{8}(-12+u^4\Phi'^2),\label{eq:tC}\\
    &\widetilde{d_+H}'+\frac{\tilde{\Sigma}'}{\tilde{\Sigma}} \widetilde{d_+H}=-\frac{\widetilde{d_+\Sigma}}{u^2\tilde{\Sigma}}\tilde{H}+\frac{\dot{h}_0}{2u^2}-\frac{\dot{h}_0\tilde{\Sigma}'}{2u\tilde{\Sigma}}+\frac{\widetilde{d_+\Sigma}\tilde{H}'}{u\tilde{\Sigma}},\label{eq:tH}\\
    &\tilde{B}''=-\frac{2}{u^3}+\frac{4u^5\alpha^8}{\tilde{\Sigma}^8}+\frac{4\widetilde{D_+\Sigma}}{u^3\tilde{\Sigma}}+\frac{\tilde{H}\dot{h}_0}{2u}+u\Phi'^2-\frac{4\widetilde{D_+\Sigma}\tilde{\Sigma'}}{u^2\tilde{\Sigma}^2}+\frac{1}{2}\dot{h}_0\tilde{H}'+\widetilde{D_+H}(\tilde{H}+u\tilde{H}'),\label{eq:tA}\\
    &\frac{\mathrm{d} \Phi'}{\mathrm{d} t}=-\frac{2\widetilde{d_+\Sigma}+(1+u\tilde{B})(-\tilde{\Sigma}+u\tilde{\Sigma}')}{u\tilde{\Sigma}}\Phi'\,.\label{eq:Phi}
\end{align}
\begin{figure}[ht]
		\begin{center}
			\includegraphics[width=0.45\textwidth]{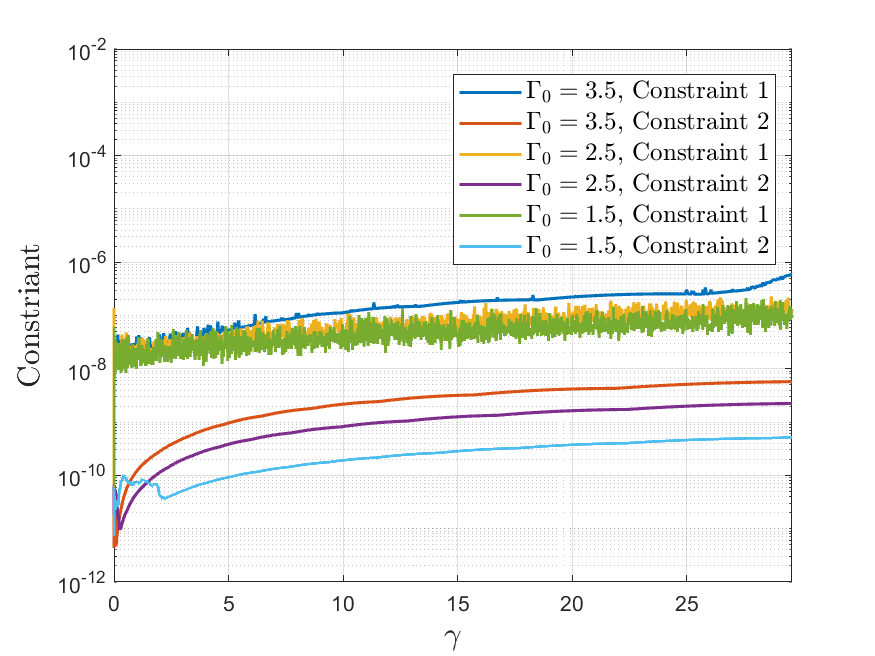}
		\includegraphics[width=0.45\textwidth]{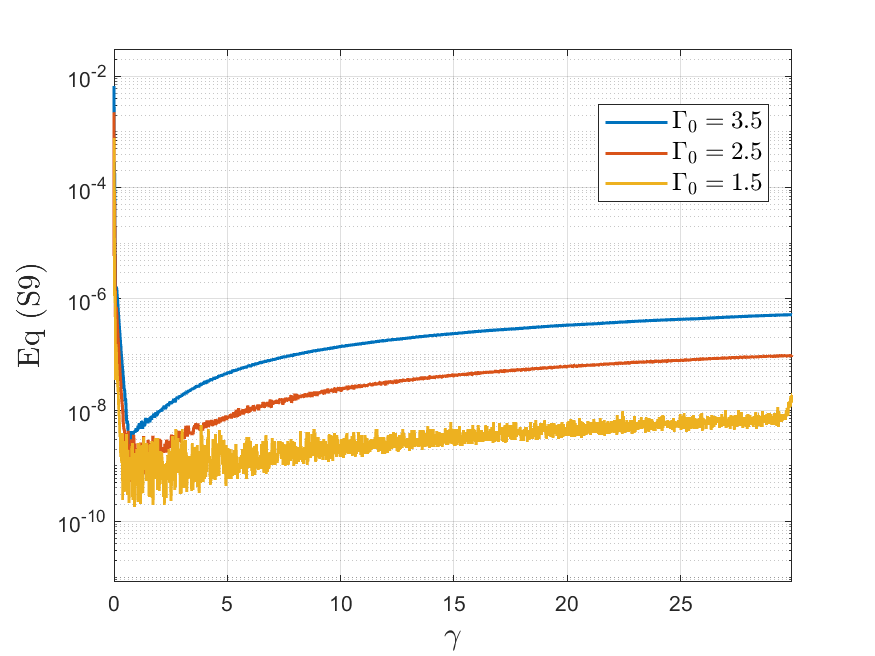}
			\caption{A numerical check of the constraints \eqref{Cons1} and \eqref{Cons2}, and of the equation \eqref{eq:boundaryB} during the time evolution for the EM model in Eq. \eqref{model1}.}
			\label{fig:cons}
		\end{center}
	\end{figure}

A schematic flow-chart of our numerical method is given as follows.
\begin{enumerate}
    \item Start from a static black hole solution with $\tilde{H}=0$ ,$\tilde{\Sigma}=s_0$ and $\Phi'=-2\rho$.
    \item Check if the constrain equations \eqref{Cons1} and \eqref{Cons2} are satisfied.
    \item Then, solve \eqref{eq:tC} for $\widetilde{d_+\Sigma}$. We use the definition of the apparent horizon to get the boundary condition for $\widetilde{d_+\Sigma}$, and fix the location of apparent horizon at $u_A=1$, $\widetilde{d_+\Sigma}=0|_{u=1}$. 
    \item Solve \eqref{eq:tH} for $\widetilde{d_+H}$. Using the  boundary asymptotic and the definition of $d_+$, we can get the boundary condition $\widetilde{d_+H}=\frac{\dot{h}_0\dot{s}_0}{s_0}+\ddot{h}_0$ at the AdS boundary.
    \item Solve \eqref{eq:tA} for $\tilde{B}$, with the boundary conditions $\tilde{B}(u=0)=\frac{2s_1-\dot{s}_0}{s_0}$ and $\tilde{B}(u=1)=-1+\frac{(d_+H)^2}{-3+\Phi'^2/4+\alpha^8/(2\Sigma^8)}|_{u=1}$. The latter is from \eqref{eq:boundaryB}. 
    \item Using the definition of $d_+$, we can get $\dot{\tilde{\Sigma}}$ , $\dot{\tilde{H}}$, and \eqref{eq:Phi} for $\dot{\Phi}$ with boundary condition $\dot{\tilde{\Sigma}}(u=0)=\dot{s}_0$, $\dot{\tilde{H}}(u=0)=\ddot{h}_0$ and $\dot{\Phi}'=-\frac{2\dot{s}_0\Phi'}{s_0}$, then we integrate in time, by employing fourth-order Runge-Kutta method for the first three time steps and then the fourth order Adams-Bashforth method, to compute $\tilde{H}(u,t+\delta t)$ and $\tilde{\Sigma}(u,t+\delta t)$, and repeat the same routine from step 2.
\end{enumerate}
In order to introduce the strain deformations avoiding large (or even infinite) initial gradients, we need to smooth the functions as follows
\begin{align}
&{\gamma}(t)=e^{-\frac{1}{a\,t+b}}\sqrt{\gamma_{1/2}\,t}\,,\\
&{\gamma}(t)=e^{-\frac{1}{at+b}}\gamma_{0}\, t\,,\\
&{\gamma}(t)=e^{-\frac{1}{at+b}}(\gamma_{2}\, t)^2\,,\\
&{\gamma}(t)=e^{-\frac{1}{at+b}}(\gamma_4\, t)^4 \,,
\end{align}
where the parameters $a$, $b$ control respectively
how fast the different kind of strain is reached and how accurate the initial configuration satisfies the constraint equation \eqref{Cons1}. We choose $a=1$ and $b=0.001$ in the numerics (here we choose the activation function of the form $e^{-\frac{1}{a\,t+b}}$; one can also use other activation functions like $\frac{1}{2}+\frac{1}{2}\tan\frac{t-t_c}{\omega_c}$, this will not affect late-time behavior of physical quantities).
As a concrete check of our numerics, in Fig. \ref{fig:cons}, we show the validity of the constraint equations \eqref{Cons1} and \eqref{Cons2}, and of equation \eqref{eq:boundaryB} during the time evolution under constant shear rates for the model of Eq.\eqref{model1}. Apart from a very short initial interval, the equations are all satisfied to a high degree of accuracy.

\end{document}